\documentstyle[12pt]{article}

\def\thefootnote{\fnsymbol{footnote}}
\def\bea{\begin{eqnarray}}
\def\eea{\end{eqnarray}}
\def\beq{\begin{equation}}
\def\eeq{\end{equation}}
\def\re{{\rm Re}}
\def\btr{{\bf Tr}}

\def\sba{\bar{\sigma}^{\dot{\beta}\alpha}}

\def\g{\bar{g}}
\def\vx{V_X}
\def\dx{\delta_X}
\def\hcD{\hat{\D}}
\def\hgg{\hat{g}}
\def\bM{\bar{M}}

\def\W{\overline{W}}
\def\cW{{\cal W}}
\def\cbW{{\cal\W}}

\def\tD{{\tilde D}}
\def\tF{{\tilde F}}
\def\D{{\cal D}}

\def\bL{\bar{\Lambda}}
\def\pp{\partial}
\def\r{\right|}
\def\l{\left.}
\newcommand{\WaWa}{\Tr({\cal W}^{\alpha}{\cal W}_{\alpha})}

\newcommand{\DbDb}{{\bar{\cal D}}^2}
\newcommand{\Da}{{\cal D}^{\alpha}}
\newcommand{\Db}{{\cal D}_{\dot{\beta}}}
\newcommand{\Dc}{{\cal D}_{\alpha}}

\def\notD{\not{\hspace{-.05in}D}}

\def\LLL{\sum_P{\lambda\Lambda^2\over 32\pi^2}}
\def\llp{\sum_P{\lambda_P\over 32\pi^2}}
\def\lll{{\Lambda^2\over 32\pi^2}}
\def\lln{{\ln\Lambda^2\over 32\pi^2}}
\def\ibar{\bar{\imath}}

\def\bmu{\bar{\mu}}
\def\bbet{\bar{\beta}}
\def\[{\left [}
\def\]{\right ]}
\def\({\left (}
\def\){\right )}
\def\lbr{\left\{}
\def\rbr{\right\}}

\def\J{\bar{J}}
\def\H{\bar{H}}

\def\T{\bar{T}}
\def\z{\bar{z}}

\def\STr{{\rm STr}}
\def\Tr{{\rm Tr}}

\def\K{{\cal K}}

\def\G{{\cal G}}
\def\f{\bar{f}}
\def\F{{\cal F}}
\def\tcF{{\tilde\F}}
\def\L{{\cal L}}
\def\bl{\bar{\lambda}}

\def\hD{\hat{D}}
\def\hG{\hat{G}}
\def\hGa{\hat{\Gamma}}
\def\hV{\hat{V}}
\def\n{\bar{n}}
\def\m{\bar{m}}
\def\s{\bar{s}}

\def\bth{\bar{\theta}}
\def\bc{\bar{\chi}}

\def\A{\bar{A}}

\def\X{{\bar{X}}}
\def\x{{\bar{x}}}
\def\Y{{\bar{Y}}}
\def\y{{\bar{y}}}
\def\Z{{\bar{Z}}}
\def\bb{\bar{b}}

\def\bv{\bar{\varphi}}
\def\hph{\hat{\varphi}}
\def\bph{\bar{\phi}}
\def\hbp{\hat{\bv}}
\def\bF{\bar{F}}
\def\bj{\bar{\jmath}}

\begin{document}

\begin{titlepage}
\begin{center}

\hfill LPTHE-Orsay-1998/46\\
\hfill LBNL-41981 \\
\hfill UCB-PTH-98/36 \\
\hfill hep-th/9806227 \\
\hfill June 1998 \\[.3in]

{\large \bf ONE-LOOP PAULI-VILLARS REGULARIZATION OF SUPERGRAVITY I:\\
CANONICAL GAUGE KINETIC ENERGY}\footnote{This work was supported in part by
the Director, Office of Energy Research, Office of High Energy and Nuclear
Physics, Division of High Energy Physics of the U.S. Department of Energy under
Contract DE-AC03-76SF00098 and in part by the National Science Foundation under
grant PHY-95-14797.} \\[.2in]

Mary K. Gaillard \\[.1in]

{\em Department of Physics and Theoretical Physics Group,
 Lawrence Berkeley Laboratory, 
 University of California, Berkeley, California 94720}\\[.5in] 

\end{center}

\begin{abstract}
%insert abstract here

It is shown that the one-loop coefficients of on-shell operators of standard 
supergravity with canonical gauge kinetic energy can be regulated by the
introduction of Pauli-Villars chiral and abelian gauge multiplets,
subject to a condition on the matter representations of the gauge group.
Aspects of the anomaly structure of these theories under global
nonlinear symmetries and an anomalous gauge symmetry are discussed.

\end{abstract}
\end{titlepage}

\newpage

\renewcommand{\thepage}{\roman{page}}
\setcounter{page}{2}
\mbox{ }

\vskip 1in

\begin{center}
{\bf Disclaimer}
\end{center}

\vskip .2in

\begin{scriptsize}
\begin{quotation}
This document was prepared as an account of work sponsored by the United
States Government. While this document is believed to contain correct 
 information, neither the United States Government nor any agency
thereof, nor The Regents of the University of California, nor any of their
employees, makes any warranty, express or implied, or assumes any legal
liability or responsibility for the accuracy, completeness, or usefulness
of any information, apparatus, product, or process disclosed, or represents
that its use would not infringe privately owned rights.  Reference herein
to any specific commercial products process, or service by its trade name,
trademark, manufacturer, or otherwise, does not necessarily constitute or
imply its endorsement, recommendation, or favoring by the United States
Government or any agency thereof, or The Regents of the University of
California.  The views and opinions of authors expressed herein do not
necessarily state or reflect those of the United States Government or any
agency thereof, or The Regents of the University of California.
\end{quotation}
\end{scriptsize}

\vskip 2in

\begin{center}
\begin{small}
{\it Lawrence Berkeley Laboratory is an equal opportunity employer.}
\end{small}
\end{center}

\newpage
\renewcommand{\theequation}{\arabic{section}.\arabic{equation}}
\renewcommand{\thepage}{\arabic{page}}
\setcounter{page}{1}
\def\thefootnote{\arabic{footnote}}
\setcounter{footnote}{0}

\section{Introduction and preliminaries}
\hspace{0.8cm}\setcounter{equation}{0}

It was shown in~\cite{pv1} that Pauli-Villars regulation of the one-loop 
quadratic divergences of a general $N=1$ supergravity theory is possible.  
This result was generalized~\cite{pv2} to the regularization of the 
one-loop logarithmic divergences of globally supersymmetric theories,
including nonlinear sigma models, with canonical kinetic energy for
Yang-Mills fields.  It was further assumed that the theory was free of
gauge and 
mixed gravitational-gauge anomalies. The purpose of the
present paper is to generalize further these results.

In section 2 we give a full PV regularization of a general supergravity
theory with canonical kinetic energy for the gauge fields and an
anomaly-free gauge group.  In section 3 we consider anomalies
under K\"ahler transformations, and in section 4 we show how the
regularization procedure must be modified in the presence of an
anomalous $U(1)$ gauge group factor. Our results are summarized in
section 5, and some calculational details, as well as 
corrections to~\cite{us,us2}, are given in appendices.

We conclude this section with a brief review of the formalism used to evaluate
the regularized Lagrangian.
The one-loop effective action $S_1$ is obtained from the term quadratic in 
quantum fields when the Lagrangian is expanded about an arbitrary background:
\bea \L_{quad}(\Phi,\Theta,c) &=& -{1\over 2}\Phi^TZ^\Phi\(\hD^2_\Phi + H_\Phi\)
\Phi + {1\over 2}\bar{\Theta}Z^\Theta\(i\notD_\Theta - M_\Theta\)\Theta 
\nonumber \\ & & + {1\over 2}\bar{c} Z^c\(\hD^2_c + H_c\)c + O(\psi), \eea
where the column vectors $\Phi,\Theta,c$ represent quantum bosons, fermions 
and ghost fields, respectively, and $\psi$ represents background fermions that 
we shall set to zero throughout this paper. The fermion sector $\Theta$ 
includes a C-odd Majorana auxiliary field
$\alpha$ that is introduced to implement the gravitino gauge fixing condition.
The full gauge fixing procedure used here is described in detail 
in~\cite{us},~\cite{us2}. The one loop bosonic action is given by
\bea S_1 &=& {i\over 2}\Tr\ln\(\hD_\Phi^2 + H_\Phi\) 
-{i\over 2}\Tr\ln\(-i\notD_\Theta + M_\Theta\) 
+ {i\over2}\STr\ln\(\hD_c^2 + H_c\) 
\nonumber \\ &=& {i\over 2}\STr\ln\(\hD^2 + H\) + T_-, \eea
where $T_-$ is the helicity-odd fermion contribution which contains no 
quadratic divergences, and the helicity-even contribution is given by
\beq \hD^2_\Theta + H_\Theta \equiv
\(-i\notD_\Theta + M_\Theta\)\(i\notD_\Theta + M_\Theta\).\eeq
The background field-dependent matrices $H(\phi)$ and $\hD_\mu(\phi) =
\pp_\mu  +
\Gamma_\mu(\phi)$ are given in~\cite{us},~\cite{us2}, where the one-loop
ultraviolet divergent contributions have been evaluated.  

We regulate the theory by including a contribution from Pauli-Villars loops,
regarded as a parameterization of the result of integrating out heavy 
({\it e.g.}, Kaluza-Klein or string) modes of an underlying finite
theory. The signature $\eta = \pm 1$ of a PV field determines the sign of its
contribution to the supertrace relative to an ordinary
particle of the same spin.  Thus $\eta = +1 (-1)$ for ordinary particles
(ghosts).  The contributions from Pauli-Villars fields with negative
signature could be interpreted as those of ghosts corresponding to heavy fields 
of higher spin.   

Explicitly evaluating (1.2) with 
an ultraviolet cut-off $\Lambda$ and a
massive Pauli-Villars sector with a squared mass matrix of the form
$$M_{PV}^2 = H^{PV}(\phi) + \pmatrix{\mu^2& \nu\cr \nu^{\dag}&\mu^2\cr} \equiv
H^{PV} + \mu^2 + \nu, \;\;\;\; |\nu|^2\sim \mu^2\gg H^{PV}\sim H, $$
gives, with $H' = H + H^{PV}$:
\bea 32\pi^2S_1 &=& - \int d^4xp^2dp^2\STr\ln\(p^2 + \mu^2 + H' + \nu\) 
+ 32\pi^2\(S'_1 + T_-\) \nonumber \\ 
&=& 32\pi^2\(S'_1 + T_-\) - \int d^4xp^2dp^2\STr\ln\(p^2 + \mu^2\) 
\nonumber \\ & &
- \int d^4xp^2dp^2\STr\ln\[1 + \(p^2 + \mu^2\)^{-1}\(H' + \nu\)\] . \eea
$S'_1$ is a logarithmically divergent contribution that involves the operator
$\hG_{\mu\nu} = [\hD_\mu,\hD_\nu]$:
\beq 32\pi^2 S'_1 = {1\over12}\int d^4xp^2dp^2\STr{1\over\(p^2 + \mu^2\)}
G'_{\mu\nu}{1\over\(p^2 + \mu^2\)}G'^{\mu\nu}, \;\;\;\; G'_{\mu\nu} = 
G_{\mu\nu} + G^{PV}_{\mu\nu}. \eeq
Finiteness of (1.4) when $\Lambda\to\infty$ requires
\bea  \STr \mu^{2n} &=& \STr H' = \STr\(2\mu^2H' + \nu^2\) = \STr \nu H'
\nonumber \\ &=& \STr H'^2 + {1\over6}\STr G'^2 + 2t'_- = 0,\eea
where $t'_-$ is the coefficient of ln$\Lambda^2/32\pi^2$ in $T_- + T^{PV}_-$.
The vanishing of $\STr \mu^{2n}$ is automatically assured by supersymmetry.
Once the remaining conditions are satisfied we obtain
\beq S_1 = - \int {d^4x\over64\pi^2}\STr\[\(2\mu^2 H' + \nu^2 + 
\STr H'^2 + {1\over6}\STr G'^2 + 2t'_-\)\ln\mu^2\].\eeq

\section{Anomaly-free supergravity}
\hspace{0.8cm}\setcounter{equation}{0}
We consider here a supergravity theory in which the Yang-Mills fields have 
canonical kinetic energy. We further assume that there are no gauge or mixed
gauge-gravitational anomalies: Tr$T^a =$ Tr$(\{T_a,T_b\}T_c) = 0$,  where $T_a$
is a generator of the gauge group. 

To regulate chiral multiplet loops, we introduce
Pauli-Villars chiral supermultiplets $Z_\alpha^I,$ % Z'^I_\alpha,$
that transform under gauge transformations like $Z^I$, $Y_I^\alpha,$
that transform according to the conjugate
representation, and gauge singlets $Y^0,Z^0$.  
Additional charged fields $X_\beta^A$ and $U^\beta_A$ transform according 
to the representation $R^a_A$ and its conjugate, respectively, under the gauge 
group factor $\G_a$, and $V^A_\beta$ transforms according to a (pseudo)real 
representation that is traceless and anomaly-free.  Their gauge couplings 
satisfy
\beq \sum_{\beta,A} \eta^A_\beta C^a_A = \sum_i C^a_i \equiv C^a_M,\eeq
where 
\beq \Tr\(T^aT^b\)_R = \delta_{ab} C^a_R \eeq
for particles transforming according to the representation $R$ (or
$\bar{R})$, and the subscripts $i,A,$ refer to the light fields and to $X,U,V,$
respectively. 
For example, if the theory has $2N_f$ fundamental representations of $\G_a$, 
(as in supersymmetric extensions of the Standard Model) we can take PV fields 
in the fundamental and anti-fundamental representations with signatures that 
satisfy $\sum_\beta\eta_\beta^f = N_f$. If there are $2N_f+1$ fundamental
representations, one needs an anomaly-free (pseudo)real representation $r$ 
for some $V^A$ such that $C^a_r = (2m + 1)C^a_f$.  If no such
representation exists, the theory cannot be regulated in this way.

To regulate gravity loops we introduce additional gauge singlets 
$\phi^\gamma$, as well as  
$U(1)$ gauge supermultiplets $W^\alpha$ with signature $\eta^\alpha$ and
chiral multiplets $Z^\alpha = e^{\theta^\alpha}$ with the same signature and 
$U(1)_\beta$ charge $q_\alpha\delta_{\alpha\beta}$, such that the
K\"ahler potential $K(\theta,\bth)= {1\over2}\nu_\alpha(\theta + \bth)^2$ is
invariant under 
$U(1)_\beta$: $\delta_\beta\theta_\alpha = -\delta_\beta\bth_\alpha = 
iq_\alpha \delta_{\alpha\beta}$.  The corresponding D-term: 
\beq \D(\theta,\bth) = \D_\theta^\alpha\D^\theta_\alpha, \;\;\;\; 
\D^\theta_\alpha = - i\sum_\beta K_\beta\delta_\alpha
\theta^\beta = q_\alpha\nu_\alpha(\theta^\alpha + \bth^\alpha),\eeq
vanishes in the background, but $(\theta^\alpha + \bth^\alpha)/\sqrt{2}$ 
acquires a squared mass $\mu_\alpha^2 = (2x)^{-1}q^2_\alpha
\nu_\alpha$ equal to that of 
$W^\alpha$, with which it forms a massive vector supermultiplet,
where $x=g^{-2}$ is the inverse squared gauge coupling, taken here to be
a constant.  

Finally, to regulate the Yang-Mills contributions, we include 
chiral multiplets $\varphi^a_\alpha,\hph^a_\alpha$ that transform 
according to the adjoint representation of the gauge group.  

We take the K\"ahler potential\footnote{This
choice is by no means unique, only illustrative.}
\bea K_{PV} &=& \sum_\gamma\[e^{\alpha^\phi_\gamma K}\phi^\gamma\bph_\gamma 
+ {1\over2}\nu_\gamma(\theta_\gamma + \bth_\gamma)^2 
+ \sum_A\(|X_\gamma^A|^2 + |U^\gamma_A|^2\)\]
\nonumber \\ & & + \sum_{\alpha,a}\(e^K\varphi_\alpha^a\bv_a^\alpha + 
\hph^a_\alpha\hbp_a^\alpha\) + 
\sum_\alpha\(K_\alpha^Z + K_\alpha^Y\) + \sum_{A\gamma}|V^A_\gamma|^2
\nonumber \\ K_\alpha^Z &=& \sum_{I,J=i,j}\[K_{i\bj}Z_\alpha^I\Z_\alpha^{\J} 
+ {b_\alpha\over2}\lbr\(K_{ij} - K_iK_j\)Z_\alpha^IZ^J_\alpha + {\rm h.c.}\rbr
\] + |Z_\alpha^0|^2, \nonumber \\ 
K_{\alpha> 3}^Y &=& \[\sum_{I,J=i,j}K^{i\bj}Y^\alpha_I\Y^\alpha_{\J} - 
a_\alpha\(Y^\alpha_I\Y_\alpha^0K^i + {\rm h.c.}\) + 
|Y^\alpha_0|^2\(1 + a^2_\alpha K^iK_i\)\] ,\nonumber \\ 
K_{\alpha\le 3}^Y &=& \sum_{I,J=i,j}\delta^{i\bj}Y_I^1\Y^1_{\J} + |Y^1_0|^2, 
\quad K^i = K^{i\m}K_{\m},\eea 
where $K^{i\m}$ is the inverse of the metric tensor $K_{i\m}$, the 
superpotential
\bea W_{PV} &=& \sum_{\alpha\beta}\[\sum_I\mu^Z_{\alpha\beta} Z^I_\alpha
 Y^\beta_I + \mu_{\alpha\beta}^0Z^0_\alpha Y^\beta_0 + 
{1\over2}\sum_a\(\mu^\varphi_{\alpha\beta}\varphi^a_\alpha\varphi^a_\beta + 
\mu^{\hph}_{\alpha\beta}\hph^a_\alpha\hph^a_\beta\)\] \nonumber \\ & & 
+ {1\over2}\sum_{\gamma}\mu^\phi_\gamma\(\phi^\gamma\)^2 
+ \sum_{A\gamma}\(\mu_\gamma^X 
U_A^\gamma X^A_\gamma + \mu_\gamma^V(V_A^\gamma)^2\) \nonumber \\ & & + 
{1\over\sqrt{2}}\sum_{\alpha=4}\(a_\alpha W_iZ^I_\alpha Y^{\alpha}_0 + 
WZ^I_\alpha Y^{\alpha}_I\) + {1\over 2} Z^I_1 Z^J_1 W_{ij} \nonumber \\ & &
+ \sqrt{{2\over x}}\sum_{\alpha=5}\varphi^a_{\alpha-4}Y_I^\alpha(T_aZ)^i 
+ {1\over2}\sum_\alpha c_\alpha|Z^0_\alpha|^2 W, \eea  
and gauge field kinetic functions
\bea f^{ab} &=& x\(\delta^{ab} +d_{\alpha\beta}\hph^a_\alpha\hph^b_\beta\), 
\quad f^{\alpha\beta} = \delta^{\alpha\beta}, \quad
f^{a\alpha} = e^{\alpha\beta}\sqrt{2x}\varphi^a_\beta, \eea
where the index $a$ refers to the light gauge degrees of freedom.
The function $K = K(Z,\Z)$ is the K\"ahler potential for 
the light chiral multiplets $Z^i = (\Z^{\ibar})^{\dag}$, $W = W(Z)$ is the
superpotential, and 
\beq   K_i = \pp_iK= {\pp\over\pp z^i}K, \quad K_{i\m} = 
\pp_i\pp_{\m}K, \quad K_{ij} = \pp_i\pp_jK, \quad etc. \eeq
Properties of the metric tensor for $Y_I,Y_0$, are given in Appendix A.  
The matrices $\mu_{\alpha\beta},d_{\alpha\beta},e_{\alpha\beta}$, are 
nonvanishing only when they couple fields of the same signature.
The parameters $\mu,\nu$, play the role of effective cut-offs; they are
constrained so as to eliminate logarithmically divergent terms of order
$\mu^2\ln\Lambda^2$ in the integral (1.4).  The parameters
$a,b,c,d,e$, are of order unity, and are chosen to satisfy:
\bea b_1 &=& 1, \quad b_{\alpha\ne 1}=0,
\quad a\equiv \sum_{\alpha=4}\eta_\alpha^Y a_\alpha^2 = -2,
\quad a'\equiv\sum_{\alpha=4}\eta^Y_\alpha a_\alpha^4= +2, 
\quad g\equiv\sum_{\alpha=5}\eta^Y_\alpha a_\alpha^2, 
\nonumber \\ 2e &\equiv& \sum_{\alpha,pv\beta}\eta^{\hph}_\alpha 
e_{\alpha\beta}^2 = 4g - 2, \quad \sum_\alpha\eta^Z_\alpha c^2_\alpha = 
- \sum_\alpha\eta^\theta_\alpha\equiv - N'_G. \eea
The signatures of the chiral PV multiplets satisfy
\bea \sum_\alpha\eta^{\varphi}_\alpha &=& 1, 
\quad \sum_\alpha\eta^{\hph}_\alpha = 2, 
\quad \sum_\alpha\eta^Z_\alpha = -1, \quad \eta^U_\alpha = \eta^X_\alpha, 
\nonumber \\ \eta^Y_\alpha  &=& \eta^Z_\alpha, \quad \eta^\varphi_\alpha =
\eta^Z_{\alpha+4}, \quad \eta^Z_1 = \eta^Z_2 = - \eta^Z_3 = -1.\eea

\subsection{Quadratic divergences}
In~\cite{pv1} it was shown how to regulate the quadratic divergences of
supergravity that are proportional to\footnote{See Appendix D for corrections
with respect to~\cite{us,us2}.  Our conventions and notations are
defined in the Appendices of these papers.}
\bea \STr H &=& -10V -2M^2 + {7\over2}r + 4K_{i\m}
\D_\mu z^i\D^\mu\z^{\m} + 2\D + N_G{r\over 2}\nonumber \\
& & + 2N\(\hV + M^2 - {r\over4}\) + 2x^{-1}\D_aD_i(T^az)^i
\nonumber \\ & & 
- 2R_{i\m}\(e^{-K}\A^iA^{\m} + \D_\nu z^i\D^\mu\z^{\m}\),\eea
where $N$ and $N_G$ are the number of chiral and gauge supermultiplets,
respectively, in the light spectrum.
In these expressions, $r$ is the space-time curvature, $R_{i\m}$ is the Ricci 
tensor associated with the K\"ahler metric $K_{i\m}$, $V = \hV
+ \D$ is the classical scalar potential with $\hV = e^{-K}A_i\A^i - 3M^2
,\;\D = (2x)^{-1}\D^a\D_a, \; \D_a = K_i(T_az)^i$,
and $M^2 = e^{-K}A\A$ is the field-dependent squared gravitino mass, with
\beq A = e^KW = \A^{\dag}, \;\;\;\; A_i = D_iA, \;\;\;\;
\A^i = K^{i\m}\A_{\m}, \;\;\;\; etc., \eeq
where $D_i$ is the scalar field reparameterization covariant derivative.

In evaluating the effective one-loop action we set to zero all background
Pauli-Villars fields; then the contribution of these fields to $\STr H$ is
\bea \STr H^{PV} &=& 2\sum_P\eta_\alpha\[{1\over x}\D^aD_P(T_az)^P -
R^P_{Pi\m}\(\A^iA^{\m}e^{-K} + \D_\mu z^i\D^\mu\z^{\m}\)\]\nonumber \\ & &
+ 2\sum_P\eta_P\(\hV + M^2\) - \(\sum_P\eta_P - \sum_\alpha 
\eta_\alpha^\theta\){r\over2}
 ,\eea
where $P$ refers to all PV chiral multiplets, including $\theta^\alpha$.
 From (2.1) we obtain for the relevant elements of the 
scalar reparameterization connection $\Gamma$ and Riemann tensor $R$
(see Appendix A):
\bea D_I(T_az_\alpha)^J &=& D_i(T_az)^j, \quad D_I(T_ay_1)^J = 
- (T_a)_j^i, \nonumber \\
(R^{Z_\alpha})^I_{Jk\m} &=& R^i_{jk\m} , \quad (R^{Y_1})^I_{Jk\m} = 0, 
\nonumber \\
D_I(T_ay_\alpha)^J &=& -D_j(T_az)^i - a^2_\alpha K_j(T_az)^i, 
\quad D_J(T_ay_\alpha)^0 = - a_\alpha(T_az)^j ,\nonumber \\
D_0(T_ay_\alpha)^I &=& a_\alpha\(K_jD_i(T_az)^j - K_{i\m}(T_az)^{\m} + 
a_\alpha^2K_i\D_a\), \quad D_0(T_ay_\alpha)^0 = a^2_\alpha\D_a,
\nonumber \\ 
(R^{Y_\alpha})^I_{Jk\m} &=& - R^j_{ik\m} - a_\alpha^2\delta^j_kK_{i\m}, 
\quad (R^{Y_\alpha})^0_{0 k\m} = a^2_\alpha K_{k\m}, \quad 
(R^{Y_\alpha})^0_{Jk\m} = 0, \nonumber \\ 
(R_\alpha^{Y_\alpha})^J_{0k\m} &=& a_\alpha\[K_iR^i_{jk\m} + 
a_\alpha^2\(K_kK_{j\m} + K_jK_{k\m}\)\], \quad \alpha > 3, \nonumber \\
D_C(T_a\phi)^D &=& (T_a)^C_D + \delta^C_D\alpha_C\D_a, \quad 
R^C_{Dk\m} = \delta^C_D\alpha_C K_{k\m}, \quad  \phi^{C,D}\ne Z,Y,\eea
where $\alpha^\varphi = 1,\; \alpha^{\hph} = \alpha^\theta = 0.$
Using these relations with (2.9) we obtain an overall contribution 
from heavy PV modes:
\bea \STr H_{PV} &=&  - {r\over2}\(N' - N'_G\) - 
2\alpha\(K_{i\m}\D_\mu z^i\D^\mu\z^{\m} - 2\D\) - 2x^{-1}\D_aD_i(T^az)^i
\nonumber \\ & & + 2\hV\(N' - \alpha\) + 2M^2\(N' - 3\alpha\) 
+ 2R_{i\m}\(e^{-K}\A^iA^{\m} + \D_\nu z^i\D^\mu\z^{\m}\),\nonumber \\
\alpha &=& \sum_C\eta_C\alpha_C , \quad N' = \sum_P\eta_P, 
\quad N'_G = \sum_\gamma\eta^\theta_\gamma .\eea
With (2.10) the finiteness condition $\STr H' = 0$ imposes the constraints 
\bea  N' &=& 7 - N ,\quad N'_G = - N_G, \quad \alpha = 2 .\eea

The vanishing of $\STr(\mu^2 H' + \nu^2)$ in (1.6)
further constrains the parameters $\mu$ and $\nu$. 
If, for example, we set\footnote{The result is unchanged if the
parameters $\mu\to\mu(z),\;\nu\to\nu(z,\z)$ depend on the light
fields\cite{pv1}.} $\mu^P_{\alpha\beta} =
\mu^P_\alpha\delta_{\alpha\beta}, \; q_a = 1,\;\mu_\alpha^{P\ne\theta}
= \beta_\alpha^{P}\mu, \; \nu_\gamma^\theta = 
\(\beta^\theta_\gamma\)^2|\mu|^2 $, the finiteness constraint requires
\bea \sum_{\alpha=1}^3\eta_\alpha^Z\(\beta^Z_\alpha\)^2 &=& 
\sum_{\alpha=4}\eta_\alpha^Z\(a_\alpha\beta^Z_\alpha\)^2 = 
N\sum_{\alpha=4}\eta_\alpha^Z\(\beta^Z_\alpha\)^2 + \sum_{C,\alpha^C=0}
\eta_C\(\beta_C\)^2 = 0, \nonumber \\ 
\sum_C\eta_C\(\beta_C\)^2 &=& 0 \quad{\rm for \; fixed} \;\alpha_C \ne
0, \quad C \ne Z^I,Y_I.\eea 
As explained in~\cite{pv1} the $O(\mu^2)$
contribution to $S_0 + S_1 = \int d^4x\(\L_0 + \L_1\)$ takes the form:
\bea \L_0(g_{\mu\nu}^0,K) + \L_1 &=& \L_0(g_{\mu\nu},K + \delta K), \;\;\;\;
g_{\mu\nu} = g_{\mu\nu}^0\(1 + \epsilon\) \nonumber \eea \bea
\epsilon &=& -\llp e^{-K}A_{PQ}\A^{PQ} = \Tr\LLL\zeta', \nonumber \\ 
\delta K &=& \llp\(e^{-K}A_{PQ}\A^{PQ} -4\K_P\) = \Tr\LLL\zeta,
\nonumber \\ \K_P &=& \delta_{P\theta_\gamma}
K^{PV}_{\theta^\gamma\bth^\gamma}\sum_\delta\delta_\delta\theta^\gamma
\delta_\delta\bth^\gamma = q_\gamma^2\nu_\gamma,\eea
where~\cite{app} 
\bea \lambda_{PQ} &=& \delta_{PQ}\lambda_P, \quad \zeta_{PQ} =
\delta_{PQ}\zeta_P, \nonumber \\ 
\lambda_P &=& 2\sum_\alpha\eta^P_\alpha
\(\beta^P_\alpha\)^2\ln\beta^P_\alpha,\;\;\;\;\zeta_{P\ne\theta} =
\zeta'_{P\ne\theta}=1,\;\;\;\;\zeta_\theta = -4,\;\;\;\;\zeta_\theta' = 0,
\nonumber \\ \(\Lambda^2\)_P^Q &=& e^KK^{Q\bar{R}}K^{\bar{T}S}\mu_{PS}
\bmu_{\T\bar{R}}, \quad
P\ne \theta,\quad \Lambda^2_{\theta_\alpha\theta_\gamma} = 
\delta_{\alpha\gamma}|\mu_\theta|^2.\eea
$\Lambda^2$ plays the role of the (matrix-valued) effective cut-off.
As emphasized previously~\cite{pv1}, if there are three or more
terms in the sum over $\alpha$, the sign of $\lambda_P$ is
indeterminate~\cite{app}.

In the following we require only on-shell invariance,\footnote{The
off-shell divergences are prescription dependent; the extension of this
regularization procedure beyond one loop may require a choice of
prescription in which they can also be made finite.} so the quadratic
divergences impose one less constraint than in (2.15).  That is, we perform a
Weyl transformation to write the one-loop corrected Lagrangian as
\bea \L_{eff} &=& \L_{tree}\(g^R\) 
- {\Lambda^2\over32\pi^2}\STr H'^2 - \epsilon\({r\over 2} + 
\D_\mu z^i\D^\mu\z^{\m}K_{i\m} - 2V\) \nonumber \\ & & 
+ O\(\lln\) + O\[\(\hbar\over16\pi^2\)^2\] + {\rm finite \; terms}, 
\nonumber \\ 
g_{\mu\nu}^R &=& (1+\epsilon)g_{\mu\nu}, \quad
\epsilon = \lll\(N + N' - N_G - N'_G - 7\),\eea
and we do not require $\epsilon$ to vanish.  Then the finiteness conditions
reduce to
\bea N' &=& 3\alpha + 1 - N, \quad N'_G = \alpha - 2 - N_G.\eea
In this case, the third finiteness condition in (1.6) becomes
\beq \STr\(2\mu^2H' + \nu^2\) = 2\STr\(\mu_G^2 - \mu^2_\chi\)
\({1\over2}r + K_{i\m}\D_\mu z^i\D^\mu\z^{\m} - 2V\) = 0.\eeq
The supertrace on the right hand side vanishes identically because the 
supertraces of the squared mass matrices $\mu^2_{PV}$ vanish separately in 
the chiral ($\mu^2_\chi$) and $U(1)$ gauge ($\mu_G^2$) PV sectors.

\subsection{Logarithmic divergences}

 From the results of~\cite{us,us2}, if $\L(g,K)$ is the standard
Lagrangian~\cite{crem,bggm} for $N=1$ supergravity coupled to
matter with space-time metric $g_{\mu\nu}$, K\"ahler potential $K$,
and gauge kinetic function $f_{ab}(Z) = \delta_{ab}$, 
the logarithmically divergent part of the one loop corrected Lagrangian is
\bea
\L_{eff} &=& \L\(g_R,K_R\) + \lln\(X^{AB}\L_A\L_B + X^A\L_A\) + \sqrt{g}\lln L
\nonumber \\ L &=& L_0 + L_1 + L_2 + L_3 + NL_\chi + N_GL_g, 
\nonumber \\ \L_A &=& {\pp \L\over\pp\phi^A},\eea
where $\phi^A$ is any light field, and\footnote{See Appendix D for corrections
with respect to~\cite{us,us2}}
\bea L_0 &=& 3C^a\delta_{ab}\(\cW^{ab} + {\rm h.c.}\) - {20\over3}\hV^2 + 
{10\over 3}\hV M^2 + 5M^4 + {88\over3}\D M^2 \nonumber \\ & & 
+ {47x\over6}\[2x\cW_{ab}\cbW^{ab} - \(F^a_{\rho\mu} - i\tF^a_{\rho\mu}\)
\(F_a^{\rho\nu} + i\tF_a^{\rho\nu}\)\D_\nu z^i\D^\mu\z^{\m}K_{i\m}\]
 \nonumber \\ & &  
- {7i\over3}\D_\mu z^i\D_\nu\z^{\m}K_{i\m}\D^aF_a^{\mu\nu}
+ {1\over3}\(25\hV + 10M^2\)K_{i\m}\D_\mu\z^{\m}\D^\mu z^i\nonumber \\ & & 
+ {20\over3}\(\cW^{ab} + \cbW^{ab}\)\D_a\D_b 
+ 11\D K_{i\m}\D_\rho z^i\D^\rho\z^{\m} \nonumber \\ & & 
- {14\over3}\D\hV + 
15\D_\mu z^j\D^\mu z^i\D_\nu\z^{\m}\D^\nu\z^{\n}K_{i\n}K_{j\m}\nonumber \\ & & 
- {20\over3}\(\D_\mu\z^{\m}\D^\mu z^iK_{i\m}\)^2 
+ {20\over 3}\D_\mu\z^{\m}\D^\mu z^i\D_\nu\z^{\n}\D^\nu z^jK_{i\n}K_{j\m},\eea 
\bea L_\chi &=& - {x\over6}\(F^a_{\rho\mu} -i\tF^a_{\rho\mu}\)
\(F_a^{\rho\nu} + i\tF_a^{\rho\nu}\)\D_\nu z^i\D^\mu\z^{\m}K_{i\m}
\nonumber \\ & &  + {1\over3}\[x^2\cW_{ab}\cbW^{ab} - 
\D\(K_{i\m}\D_\rho z^i\D^\rho\z^{\m} + 2\hV + 4 M^2\)\]\nonumber \\ & & 
+ {1\over 3}\(\hV + 2M^2\) K_{i\m} \D_\mu\z^{\m}\D^\mu z^i 
- {i\over3}\D_\mu z^j\D_\nu\z^{\m}K_{i\m}\D^aF_a^{\mu\nu}\nonumber \\ & & 
+ {2\over 3}\hV M^2 + M^4  
+ {1\over3}\D_\mu z^j\D^\mu z^i\D_\nu\z^{\m}\D^\nu\z^{\n}K_{i\n}K_{j\m} , \eea 
\bea L_1 &=& -\[\cW^{ab}D_i(T_bz)^jD_j(T_az)^i + {\rm h.c.}\] 
+ {2\over x}\D_\mu z^i\D^\mu\z^{\m} R^k_{i\m j}\D_a D_k(T^a\z)^j 
\nonumber \\ & &  + {2\over x}\D_ae^{-K}R^{k\;j}_{\;n\;i}A_k\A^nD_j(T^az)^i 
+ 2iF^a_{\mu\nu}D_j(T_az)^iR^j_{i\m k} \D^\mu z^k\D^\nu\z^{\m} 
\nonumber \\ & &
+ \D_\mu z^j\D^\mu\z^{\m}R^k_{j\m i}\D_\nu z^{\ell}\D^\nu\z^{\n}R^i_{\ell\n k}
+ \D_\mu z^j\D_\nu\z^{\m}R^k_{i\m j}
\D^\mu z^{\ell}\D^\nu\z^{\n}R^i_{k\n \ell}\nonumber \\ & & 
- \D_\mu z^j\D_\nu\z^{\m}R^k_{i\m j}
\D^\nu z^{\ell}\D^\mu\z^{\n}R^i_{k\n \ell}
+ 2e^{-K}\D_\mu z^i\D^\mu\z^{\m}R^k_{i\m j}R^{\ell\;j}_{\;n\; k}A_{\ell}\A^n
\nonumber \\ & &  + e^{-2K} A_i\A^kR^{m\;i}_{\;n\;k}R^{n\;p}_{\;m\;q}A_p\A^q, 
\eea  \bea L_2 &=& {2\over3x}D_i(T_az)^i\D_a\(\D_\mu z^j\D^\mu\z^{\m}K_{j\m} 
+ \hV + 3M^2\) + {2i\over3}\D_\mu z^i\D_\nu\z^{\m}R_{i\m}\D^aF_a^{\mu\nu} 
\nonumber \\ & & + {2\over3}D_i(T_az)^i\[\(\cW^{ab} + \cbW^{ab}\)\D_b 
+ ixF^a_{\mu\nu}K_{\m j}\D^\mu z^j\D^\nu\z^{\m}\] 
\nonumber \\ & & + {4\over 3}\D e^{-K}R^i_jA_i\A^j
- {2\over 3} \D_\mu z^i\D^\mu\z^{\m}\[e^{-K}R^k_nA_k\A^n K_{i\m}
+ R_{i\m}\(\hV + 3M^2\)\] \nonumber \\ & & 
- {2\over3}\D_\mu z^i\D_\nu\z^{\m}K_{i\m}R_{j\n}\(\D^\mu z^j\D^\nu\z^{\n}
- \D^\nu z^j\D^\mu\z^{\n}\) - {2\over 3}e^{-2K}R^m_nA_m\A^nA_j\A^j 
\nonumber \\ & & 
- {2\over 3}\D_\rho z^i\D^\rho\z^{\m}K_{i\m}\D^\mu z^j\D_\mu\z^{\n}R_{j\n}
+ {4\over3}\D\D_\mu z^i\D^\mu\z^{\m}R_{i\m} , 
\eea \bea L_3 &=& \D_\mu z^j\D^\mu z^iR^{k\;\;\ell}_{\;\;j\;\; i}
\D_\nu\z^{\n}\D^\nu\z^{\m}R_{\n k\m\ell}  \nonumber \\ & & 
+ e^{-K}\[\D_\mu z^i\D^\mu z^j\( A_{ik\ell}\A^{n}R^{k\;\ell}_{\;n\;j}
- R^{k\;\ell}_{\;j\; i}(A_{mk\ell}\A^m -A_{k\ell}\A)\) + {\rm h.c.}\]
\nonumber \\ & & + {e^{-K}\over x}\D_a\[(T^az)^iR_{i\;\;\ell}^{\;\; j\;\; k}
\A^{\ell}A_{jk} + {\rm h.c.}\]
+ e^{-2K}\(R^{j\;k}_{\;n\;i}A_{jk}\A^nA\A^i + {\rm h.c.}\)\nonumber \\ & &
+ e^{-K}\(2\D_\mu z^i\D^\mu\z^{\m} + e^{-K}\A^iA^{\m}\)R^{\ell}_{j\m k}
R^{j\;k}_{\;i\;n}A_{\ell}\A^n \nonumber \\ & &
- \(\D_\mu z^i\D^\mu\z^{\m} + e^{-K}\A^iA^{\m}\)\[\D_i\(e^{-K}R^k_{\ell\m j}
A_k\A^{j\ell}\) + {\rm h.c.} \], \eea 
\bea L_g &=& {1\over3}K_{i\m}K_{j\n}\(2\D_\mu z^i\D^\mu z^j\D_\nu\z^{\m}
\D^\nu\z^{\n} + \D_\mu z^i\D^\mu\z^{\n}\D_\nu\z^{\m}\D^\nu z^j\)
\nonumber \\ & & -{1\over3}\(\D_\mu z^i\D^\mu\z^{\m}K_{i\m}\)^2
+ x^2\cW_{ab}\cbW^{ab} + {1\over3}\(\cW_{ab} + \cbW_{ab}\)\D^a\D^b
\nonumber \\ & & 
- {1\over3}\hV^2 + {1\over3}\(\hV + \D\)\D_\mu z^i\D^\mu\z^{\m}K_{i\m} 
- {i\over3}\D_\mu z^i\D_\nu\z^{\m}K_{i\m}\D^aF_a^{\mu\nu} \nonumber \\ & & 
- {2\over3}\hV\D - {x\over2}K_{i\m}\D_\nu z^i\D^\mu\z^{\m}\(F^a_{\rho\mu} + 
i\tF^a_{\rho\mu}\)\(F_a^{\rho\nu} - i\tF_a^{\rho\nu}\), \eea 
where $F^2 = F^a_{\mu\nu}
F_a^{\mu\nu}$ with $F^a_{\mu\nu}$ the Yang-Mills field strength, 
\beq \cW_{ab} = {1\over4}\(F_a\cdot F_b + \tF_a\cdot F_b\)
- {1\over2x}\D_a\D_b,\eeq
and
\bea e^KD_i\(e^{-K}R^\ell_{j\m k}A_\ell\A^{jk}\) &=& 
(D_iR^{\ell}_{j\m k})A_{\ell}\A^{jk} + R^{\ell}_{j\m k}A_{i\ell}\A^{jk} 
\nonumber \\ & & + 2R^k_{i\m j}A_k\A^j + 
R^{\ell}_{j\m k}R^{j\;k}_{\;i\;n}A_{\ell}\A^n . \eea
The renormalized K\"ahler potential is
\bea K_R &=& K + \lln\[e^{-K}A_{ij}\A^{ij} -2\hV - 10M^2 - 4\K^a_a -12\D\],
\nonumber \\ \K^a_b &=& {1\over x}(T^az)^i(T_b\z)^{\m}K_{i\m}. \eea
The second term in the expression (2.24) for $\L_{eff}$ does not contribute 
to the S-matrix. Since we are only interested in on-shell finiteness, we can 
drop it.  We have also dropped total derivatives, including the Gauss-Bonnet 
term which can readily be extracted from the results of~\cite{us,us2}:
\beq \L_{eff}\ni \sqrt{g}\lln{1\over48}\(41 + N -3N_G\)
 \(r^{\mu\nu\rho\sigma}r_{\mu\nu\rho\sigma} - 4r^{\mu\nu}r_{\mu\nu} + r^2\), 
\eeq
in agreement with other calculations~\cite{GB}.  We similarly drop total
derivatives in the logarithmically divergent PV contributions.

The Pauli-Villars contribution to (2.24) is, after an appropriate additional
space-time metric redefinition,
\bea \L_{PV} &=& \sqrt{g}\lln\[N'_G L_g + N' L_\chi + \sum_P\eta_P\(L_1^P + 
L_2^P\) + L_3^Z + L_{\cW} + eL_e\] \nonumber \\ & & \quad + 
\Delta_{K'}\L,\quad \quad K' = \lln e^{-K}\sum_{P,Q}\eta_PA_{PQ}\A^{PQ},
	\label{lpv}\eea where 
\bea {1\over\sqrt{g}}\Delta_F\L &=& \Delta_FL = - F\hV + 
\(e^{-K}\A^iA^{\m} + \D_\mu z^i\D^\mu\z^{\m}\)\pp_i\pp_{\m}F \nonumber \\ & &
- \lbr \pp_iF\[e^{-K}\A^i A + {1\over2x}\D_a(T^az)^i\] + {\rm h.c.}\rbr,
\label{dlk}\eea
is the shift in $\L/\sqrt{g}$ due to a shift $F(z,\z)$ in the
K\"ahler potential, and [see Appendix B and Eq. (B.38)]
\bea L_{\cW} &=& x^2\cW_{ab}\cbW^{ab}\[2e^2 + (d-2e)^2\], \nonumber \\ 
L_e &=& 2i\D_\mu z^j\D_\nu\z^{\m}K_{i\m}\D^aF_a^{\mu\nu} + 4\D\(3M^2 + \hV\) 
- 4x^2\cW_{ab}\cbW^{ab} \nonumber \\ & &  
+ x\(F^a_{\rho\mu} -i\tF^a_{\rho\mu}\)
\(F_a^{\rho\nu} + i\tF_a^{\rho\nu}\)\D_\nu z^i\D^\mu\z^{\m}K_{i\m}
 \nonumber \\ & & + 2\D\D_\mu z^i\D^\mu\z^{\m}K_{i\m}   - 4\Delta_{\D}L, 
\label{lw}\eea 
are the contributions from the gauge kinetic terms given in (2.6),
obtained by a 
straightforward generalization of the results of~\cite{us2} to the case of a
nondiagonal gauge kinetic function $f_{ab}$ (see Appendix B).  

To evaluate $K'$ and $L_3$ we need the additional PV matrix elements
(see appendix A):
\bea R^{Z_1}_{I\m J\n} &=& R_{i\m j\n} + K_{i\m}K_{j\n} + K_{i\m}K_{j\n},\quad
A^{Z_1}_{IJ} = A_{ij}, \quad \A_{Z_1}^{IJ} = \A^{ij}, \nonumber \\
A^{Y_\alpha,\varphi_{\alpha-4}}_{Ia} &=& e^K\sqrt{{2\over x}}(T_az)^i, \quad 
\A_{Y_\alpha,\varphi_{\alpha-4}}^{Ia} = \sqrt{{2\over x}}\[(T^a\z)^{\m}
K_{i\m} + a_\alpha^2 K_i\D^a\],\nonumber \\
\sqrt{2}A^{Z_\alpha,Y_\alpha}_{IJ} &=& \sqrt{2}A\delta^j_i, \quad 
\sqrt{2}\A_{Z_\alpha,Y_\alpha}^{IJ} = \delta^i_j\A + a^2_\alpha
e^KK_jW\A^i, \quad \alpha > 3, \nonumber \\ 
\sqrt{2}\A_{Z_\alpha,Y_\alpha}^{I0} &=& a_\alpha \A^i, \quad 
\sqrt{2}A^{Z_\alpha,Y_\alpha}_{I0} = a_\alpha e^KW_i, \quad \alpha > 3, 
\nonumber \\ A^\theta_{\alpha\beta} &=& \delta_{\alpha\beta}\nu_\alpha W,
\quad \A^{\alpha\beta}_\theta = \delta^{\alpha\beta}\nu^{-1}_\alpha\W,
\nonumber \\ A^{Z^0}_{\alpha\beta} &=& \delta_{\alpha\beta}c_\alpha W,
\quad \A^{\alpha\beta}_{Z^0} = \delta^{\alpha\beta}c_\alpha\W,\eea
where we have not included $\mu$-dependent terms that are already contained in
(2.17).  Then, using (2.8-9) we obtain
\beq K' = - \lln\[e^{-K}A_{ij}\A^{ij} + 2\hV + 2M^2 - 4\K^a_a - 
4(e + 1)\D\].\eeq
$L_3$ is determined by the expressions
\bea R^{Z_1}_{I\m J\n}\(R^{Z_1}\)^{I\;\; J}_{\;k\;\;\ell} &=& 
R_{i\m j\n}R^{i\;\; j}_{\;k\;\;\ell} + 4R_{k\m\ell\n}  + 2\(K_{k\m}K_{\ell\n} 
+ K_{\ell\m}K_{k\n}\), \nonumber \\
A^{Z_1}_{IJ}\A_{Z_1}^{IJ} &=& A_{ij}\A^{ij}, \quad
R^{Z_1}_{I\m J\n}\A_{Z_1}^{IJ} = R_{i\m j\n}\A^{ij} + 2\A_{\m\n}, \eea
giving
\bea
L_3^Z &=& - L_3 + 4\Delta_{\hV}L + 8\Delta_{M^2}L - {2\over\sqrt{g}}e^{-K}
\(A_i\A\L^i +  {\rm h.c.}\) \nonumber \\ & &
- 4\D_\mu z^j\D^\mu z^i\D_\nu\z^{\m}\D^\nu\z^{\n}\(K_{i\n}K_{j\m} +
R_{i\m j\n}\) - 4M^2\(2\hV + 3M^2\) \nonumber \\ & &
- 4e^{-K}\(2\D_\mu z^i\D^\mu\z^{\m} + e^{-K}\A^iA^{\m}\)R^{\ell}_{i\m n}
A_{\ell}\A^n - 8\D M^2,\label{l3z}\eea
where relations among operators given in Appendix B of~\cite{us2} were used.
$L_2^P$ is obtained directly from (2.13):
\bea \sum_P\eta_PL_2^P &=& - L_2 - {2\over3}\alpha L_\alpha, \nonumber \\ 
L_\alpha &=& \(\hV + 3M^2\)^2 - 
4\D\(K_{i\m}\D_\rho z^i\D^\rho\z^{\m} + \hV + 3 M^2 \)
\nonumber \\ & & + 2\(\hV + 3M^2\)K_{i\m}\D_\mu\z^{\m}\D^\mu z^iK_{i\m} 
+ \(\D_\mu\z^{\m}\D^\mu z^iK_{i\m}\)^2 \nonumber \\ & & 
+ \D^\mu z^i\D^\nu\z^{\n}K_{i\n}K_{j\m}\(\D_\mu z^j\D_\nu\z^{\m}
- \D_\mu\z^{\m}\D_\nu z^j\) \nonumber \\ & & 
- 2i\D_\mu z^j\D_\nu\z^{\m}K_{i\m}\D^aF_a^{\mu\nu} 
- \(\cW^{ab} + \cbW^{ab}\)\D_a\D_b  .\label{l2}\eea 
To evaluate $L_1^P$ we need
\bea D_I(T_az_\alpha)^JD_J(T_bz_\alpha)^I &=& D_i(T_az)^jD_j(T_bz)^i, 
\nonumber \\
D_I(T_ay_1)^JD_J(T_by_1)^I &=& - \delta_{ab}C^a_M , 
\nonumber \\
(R^{Z_\alpha})^I_{Jk\m}(R^{Z_\alpha})^J_{I\ell\n} &=& R^i_{jk\m}R^j_{i\ell\n} , 
\nonumber \\ 
(R^{Z_\alpha})^J_{Ik\m}D_J(T_bz_\alpha)^I &=& R^i_{jk\m}D_j(T_bz)^i, 
\nonumber \\
D_P(T_ay_\alpha)^QD_Q(T_by_\alpha)^P &=& D_j(T_az)^iD_i(T_bz)^j + a^2_\alpha 
x\(\K_{ab} + \K_{ba}\), \nonumber \\
(R^{Y_\alpha})^P_{Qk\m}(R^{Y_\alpha})^Q_{P\ell\n} &=& R^j_{ik\m}R^i_{j\ell\n} 
- 2a^2_\alpha R_{\ell\n k\m} + a^4_\alpha\(K_{k\m}K_{\ell\n} +
K_{k\n}K_{\ell\m}\), \nonumber \\ 
(R^{Y_\alpha})^P_{Qk\m}D_Q(T_by^\alpha)^P &=& R^j_{ik\m}D_i(T_bz)^j + 
a^2_\alpha D_k(T_bz)^jK_{j\m}, \quad \alpha > 3, \nonumber \\
D_C(T_a\phi)^DD_D(T_b\phi)^C  &=& C^a_C + \delta^C_D\alpha^2_C\D_a\D_b, \quad 
R^C_{Dk\m}R^D_{C\ell\n} = \delta^C_D\alpha^2_C K_{k\m}K_{\ell\n}, \nonumber \\
R^C_{Dk\m}D_D(T_b\phi)^C &=& \delta^C_D\alpha^2_C K_{k\m}\D_b,
 \quad  \phi^{C,D}\ne Z,Y.\eea  
Then using the constraints (2.8) and the results given in Appendix B
of~\cite{us}, we obtain (see Appendix A)
\bea \sum_P\eta^PL_1^P &=& - L_1 - 3C^a\delta_{ab}\(\cW^{ab} + {\rm h.c.}\) + 
\alpha' L_\alpha + L_1^Y, \quad \alpha' = \sum_C\eta_C\alpha^2_C , \nonumber \\ 
L_1^Y &=& 4\[\Delta_{M^2}\L + M^2\(2\hV + 3M^2 + 2\D\)\]\nonumber \\ & & 
+ 8\Delta_{\D}L -
{2\over x\sqrt{g}}\[\D_a(T^az)^i\L_i + i\D_\mu\z^{\m}(T^az)^i
K_{i\m}\L^\mu_a + {\rm h.c.}\] \nonumber \\ & & 
+ 4\D_\mu z^j\D^\mu z^i\D_\nu\z^{\m}\D^\nu\z^{\n}\(R_{i\m j\n} + 
K_{i\n}K_{j\m} \) \nonumber \\ & &
+ 4e^{-K}\(2\D_\mu z^i\D^\mu\z^{\m} + e^{-K}\A^iA^{\m}\)R^{\ell}_{i\m n}
A_{\ell}\A^n .\label{l1}\eea
Adding the above, we get for the total PV contribution:
\bea
\L_{PV} &=& \lln\(X_{PV}^{AB}\L_A\L_B + X_{PV}^A\L_A\) + \sqrt{g}\lln L_{PV}
+ \Delta_{K^{PV}}\L, \nonumber \\ 
K^{PV} &=& K' + \lln\[4\hV + 12M^2 + 8\D\] = - \(K_R - K\),
\nonumber \\ L_{PV} &=& N'_G L_g + N' L_\chi - L_1 - L_2 - L_3  + L_{\cW} 
+ eL_e \nonumber \\ & & 
+ \(\alpha' - {2\over3}\alpha\)L_\alpha.\eea
The renormalization of the K\"ahler potential is seen to be finite. Setting
\beq 2e^2 + (d-2e)^2 = 2e, \eeq and using the
constraints (2.20), we obtain for the remaining contributions
\bea L + L_{PV} &=& -\(6 + \alpha - \alpha'\)\[\hV^2 +
\D_\mu\z^{\m}\D^\mu z^i\D_\nu\z^{\n}\D^\nu z^j\(K_{i\m}K_{j\n} -
K_{i\n}K_{j\m}\)\] \nonumber \\ & & \(2 -\alpha + 3\alpha'\)\(2\hV M^2 + 3M^4  
+ 2M^2 K_{i\m}\D_\mu\z^{\m}\D^\mu z^iK_{i\m} \)
\nonumber \\ & & + 2\(4 + \alpha' \)\hV K_{i\m}\D_\mu\z^{\m}\D^\mu z^iK_{i\m} 
\nonumber \\ & & + \(14 + \alpha + \alpha'\)
\D_\mu z^j\D^\mu z^i\D_\nu\z^{\m}\D^\nu\z^{\n}K_{i\n}K_{j\m}\nonumber \\ & & 
+ 4\(7 + \alpha - 3\alpha' + 3e\)\D M^2 \nonumber \\ & & 
+ \(6 + \alpha - \alpha'\)\(\cW^{ab} + \cbW^{ab}\)\D_a\D_b 
\nonumber \\ & & + 2\(7 + \alpha - e\)x\Bigg[x\cW_{ab}\cbW^{ab} \nonumber \\
& & \quad \quad - {1\over2}
\(F^a_{\rho\mu} - i\tF^a_{\rho\mu}\)\(F_a^{\rho\nu} + i\tF_a^{\rho\nu}\)
\D_\nu z^i\D^\mu\z^{\m}K_{i\m}\Bigg]  \nonumber \\ & & 
- 2\(1 + \alpha' - e\)\(2\D\hV + i\D_\mu z^j\D_\nu\z^{\m}
K_{i\m}\D^aF_a^{\mu\nu}\)\nonumber \\ & & + 2\(5 + \alpha - 2\alpha' + e\)
\D K_{i\m}\D_\rho z^i\D^\rho\z^{\m} .\eea
Finiteness is achieved by imposing
\beq \alpha = -10, \quad \alpha' = -4, \quad e = -3.\eeq

Once all the infinities have been removed, the Lagrangian takes the form (1.7),
with the matrix-valued effective cut-off a function of the scalar
fields. In particular, the terms of order $\ln\mu$ are given by (2.22)
with $\ln\Lambda^2$ replaced by the matrix $\sum_P\eta^P\ln(\mu^2_P).$

\section{K\"ahler anomalies}
\hspace{0.8cm}\setcounter{equation}{0}

Classically, supergravity theories are invariant 
K\"ahler transformations that redefine the K\"ahler potential and the
superpotential in terms of a holomorphic function $H(z)$:
\beq K\to K+ H+\H,\quad W\to e^HW,\label{kahler}\eeq
and that shifts the the fermion axial $U(1)$ current:
\beq \Gamma_\mu = {i\over4}\(\D^\mu z^iK_i - \D^\mu\z^{\m}K_{\m}\)\to
\Gamma_\mu  - {1\over2}\pp_\mu{\rm Im}H.\label{chiral}\eeq
This invariance is anomalous at the quantum level due to the conformal
and chiral anomalies.
Consider for example the one-loop correction to the Yang-Mills term:
\bea \L_1^{YM} &=& - {1\over 16\pi^2}\({1\over4}F^{\mu\nu}_aF_{\mu\nu}^a - 
{1\over2x}\D_a\D^a\)\sum_P\eta_PC^a_P\ln\(\Lambda^2_P\beta^2_P\) + \cdots, 
\nonumber \\ &=& - {1\over16\pi^2}\({1\over4}F^{\mu\nu}_aF_{\mu\nu}^a - 
{1\over2x}\D_a\D^a\)\[3C^a_G\ln\(e^{K/3}\mu^2_\varphi\rho_\varphi\) 
- C^a_M\ln\(e^K\mu_Z^2\rho_Z\)\]\nonumber \\ & &\quad + \cdots, \label{ym}\eea
in the notation of (2.16), where the dots represent operators of higher
dimension, and~\cite{app}
\beq \ln\rho_{\varphi} =  \sum_{\alpha,P=\varphi,\hph}\ln\(\beta^P_\alpha\)^2, 
\quad \ln\rho_Z = \sum_{\alpha,P=Z,X,V}\ln\(\beta^P_\alpha\)^2.\eeq
Under (\ref{kahler}) the quantum correction (\ref{ym}) changes by
\bea \delta\L_1^{YM} &=& - {{\rm Re}H\over8\pi^2}
\({1\over4}F^{\mu\nu}_aF_{\mu\nu}^a - {1\over2x}\D_a\D^a\)\(C^a_G - C^a_M\),
\label{dym}\eea
Gauginos and chiral fermions have K\"ahler $U(1)$ weights $+1$ and $-1$,
respectively, so the corresponding chiral anomaly
\bea \delta_\chi\L_1^{YM} &=& - {{\rm Im}H\over8\pi^2}\({1\over4}F^{\mu\nu}_a
\tF_{\mu\nu}^a - {1\over2x}\D_a\D^a\)\(C^a_G - C^a_M\).\label{cdym}\eea
combines with (\ref{dym}) to give the superfield expression
\bea \delta\L_1^{YM} &=& - {1\over8\pi^2}\int d^4\theta{E\over8R}W^\alpha_a
W_\alpha^a\(C^a_G - C^a_M\).\label{sdym}\eea
The field dependence of the effective cut-offs was in fact determined 
in~\cite{tom} by imposing the supersymmetric relation 
between the chiral and conformal anomalies associated with K\"ahler
transformations; this in turn restricts the K\"ahler potential for charged 
PV fields.

Sigma-models coupled to supergravity are invariant under a group of
nonlinear transformations $Z \to f(Z)$ that effect a K\"ahler transformation
of the form (\ref{kahler}), (\ref{chiral}).  
This is in general a classical invariance, and an 
interesting question is under what circumstances this invariance, which
we will refer to 
as modular invariance, can be respected at the quantum level.  If modular 
invariance is broken at the quantum level, the resulting chiral and conformal
modular anomalies must form a supermultiplet.  We consider some examples
below.

\subsection{Nonlinear sigma-models}

Consider first an ungauged supergravity theory with no
superpotential and with a K\"ahler metric typically of the form
\beq K = \sum_{A=1}^m K^A, \;\;\;\; K^A = -{1\over k_A}\ln\(1 + 
\eta\sum_{i=1}^{n_A}|z_A^i|^2\), \quad k_A = - \eta|k_A|\label{mod}, \eeq
that is classically invariant under the infinitesimal nonlinear 
transformations 
\beq \delta z^i_A = \beta^i_A + \eta z^i_A\sum_j\bbet^{\bj}_Az^j_A, \quad
\delta K^A = F^A + \bF^A, \quad F^A = \sum_j\bbet^{\bj}_Az^j_A,\label{sig}\eeq
where $\eta = + (-) 1$ for a (non)compact symmetry group.  Then the derivatives
of the metric satisfy
\bea  K^A_{jk} &=& k_AK^A_jK^A_k, \;\;\;\; 
\Gamma^{Ai}_{jk} = k_A\(\delta^{Ai}_jK^A_k 
+ \delta^{Ai}_kK^A_j\), \nonumber \\
R^{Ai}_{jk\m} &=& k_A\(\delta^{Ai}_jK^A_{k\m} 
+ \delta^{Ai}_kK^A_{j\m}\), \;\;\;\; \delta^{Ai}_j = 
\cases{\delta^i_j & if $K^A_i \ne0$ \cr0&if $K^A_i =0$\cr}.\label{metric}\eea

To regulate the theory, we need only include a subset of the chiral
supermultiplets in (2.4). We take the K\"ahler potential
\bea K_{PV} &=& \sum_\gamma e^{\sum_A\alpha^A_\gamma K^A}
\phi^\gamma\bph_\gamma  + \sum_{A,\alpha}
K_{A,\alpha}^Z + K_{A,\alpha}^Y,\nonumber \\ 
K_{A,\alpha}^Z &=& \sum_{I,J=i,j}\[K^A_{i\bj}Z_{A,\alpha}^I\Z_{A,\alpha}^{\J}
+{b_\alpha\over2}\(K^A_iK^A_jZ_{A,\alpha}^IZ^J_{A,\alpha} 
+ {\rm h.c.}\)\], \nonumber \\
K_{A,\alpha}^Y &=& e^{\sum_B\alpha^B_{A\alpha}}K\sum_{I,J=i,j}
K_A^{i\bj}Y_I^{A,\alpha}\Y^{A,\alpha}_{\J}, 
\quad \eta_{A,\alpha}^Z =  \eta_{A,\alpha}^Y \equiv
\eta^A_\alpha,\label{skp} \eea 
and the superpotential
\bea W_{PV} &=& \sum_{I,A,\alpha\beta}\mu^Z_{A,\alpha\beta} Z^I_{A,\alpha}
Y^{A,\beta}_I +
\sum_{\alpha\beta}\mu^\phi_{\alpha\beta}\phi^\gamma\varphi^\beta,\label{sw}\eea
where $ \mu_{\alpha\beta} = 0$ if $\eta_\alpha \ne \eta_\beta$.

Then (2.10) and (2.12) reduce to
\bea \STr H &=& 2\sum_A\D_\mu z^i\D^\mu\z^{\m}K^A_{i\m}\[2 - k_A\(n_A + 1\)\] +
{r\over2}(7 - N), \nonumber \\
\STr H_{PV} &=& - 2\sum_A\alpha_A\D_\mu z^i\D^\mu\z^{\m}K^A_{i\m} -
{r\over2}N', \nonumber \\ N &=& \sum_An_A, \quad N' = 
\sum_\gamma \eta_\gamma + 2n_A\sum_{\alpha,A}\eta^A_\alpha,
\nonumber \\ \alpha_A &=& \sum_\alpha\eta^\phi_\alpha\alpha^A_\alpha + 
\sum_B\eta^B_\alpha\alpha^A_{B\alpha}.\eea 
Cancellation of the on-shell quadratic divergences requires 
\beq N + N' = 2\alpha_A + 2k_A(n_A + 1) + 3, \eeq
and additional constraints on the parameters provide a cancellation of all
one-loop ultraviolet divergences.

The PV K\"ahler potential (\ref{skp}) is invariant under the K\"ahler
transformation (\ref{mod}), provided the PV superfields transform as
\bea \delta Z^I_A &=& {\pp\delta z^i_A\over z^j_A}Z^J_A = \eta\(Z^I_A
F_A + z^i_A\sum_j\bbet^{\bj}_AZ^J_A\), \quad 
\delta\phi^\alpha = -\sum_A\alpha_\alpha^AF^A\phi^\alpha, \nonumber \\
\delta Y^A_I &=& = - \eta\(Y^A_IF_A + \bbet^{\ibar}_A\sum_jz^j_AY^A_J\)
- Y_I^A\sum_B\alpha^B_AF^B.\eea
To obtain a fully invariant PV potential requires 
\beq \alpha^A_{B\alpha} = 1,\quad \mu^\phi_{\alpha\beta} = 0 \;\; {\rm
if} \;\; \alpha^\phi_\alpha + \alpha^\phi_\beta \ne 1,\eeq
in which case the superpotential (\ref{sw}) transforms under 
(\ref{mod}) as $\delta W_{PV} = - W_{PV}\sum_AF^A,$ and
the effective cut-offs $\Lambda^2_{PQ}$ are constant. However in this case 
\beq \alpha_A = {1\over2}N',\quad H_{PV} = -N'\(\D_\mu z^i\D^\mu\z^{\m}K_{i\m} 
+ {r\over2}\), \eeq
which is removed by the Weyl transformation (2.19). Thus chiral
supermultiplets with modular invariant masses do not contribute to
quadratic divergences, nor do massive abelian gauge multiplets.   Since
modular invariance of their masses requires $\alpha^\theta$ = 0,
$\theta$-loops contribute only to the space-time curvature term and
exactly cancel the corresponding gauge loop contributions.  Therefore,
modular invariant regularization cannot be achieved unless the massless
theory is free of quadratic divergences.  This requires a constraint on
the total massless spectrum.  If it includes $N_G$ gauge supermultiplets
and $N_q$ additional chiral supermultiplets $\phi^\alpha$ with modular
weights $q^A_\alpha$, that is, with K\"ahler potential
\beq K(\phi^\alpha,\bph^\alpha) = \sum_\alpha|\phi^\alpha|^2
e^{\sum_Aq^A_\alpha K^A},\eeq
the constraint reads 
\beq 2\sum_\alpha q^A_\alpha - N_q - N + N_G + 3 + k_A(n_A + 1) = 0.\eeq
If this constraint is satisfied, the K\"ahler potential is not renormalized,
and the 
classical Bagger-Witten quantization condition~\cite{sjr,bag}, which relates 
the pion decay constant to the Planck mass in a compact $\sigma$-model,
is preserved at the quantum level.  If this is not the case, one can
still preserve the BW condition by imposing, in addition to (2.16), the
additional constraints [see (2.17-18)] on the PV masses:
\beq \sum_{\alpha\beta}\eta_\alpha\beta^2_{\alpha\beta}
\ln\(\beta_{\alpha\beta}\) = 0 \;\; {\rm for\; fixed}\;\; 
\alpha_\alpha + \alpha_\beta\ne 1.\label{masses}\eeq

If the group of modular transformations is noncompact, a subgroup of the
modular transformations (\ref{sig}) may be a classical invariance of the Lagrangian
in the presence of a superpotential and of gauge interactions for a
subset of the $Z^i$.  An example is the Lagrangian for the 
``untwisted sector'' of light fields in a class
of orbifold compactifications of the heterotic string.  The K\"ahler potential
is (neglecting the dilaton) 
\beq K = \sum_{I=1}^3G^I, \quad
G^I = - \ln\(T_I + \T_I - \sum_{A=1}^{n-1}|\Phi^A_I|^2\).\label{noscale}\eeq
It is invariant under an $SL(2,R)$ group of modular transformations that leave
$K$ invariant, and the derivatives of $K$ satisfy (\ref{metric}) with 
$K^A\to G^I,\;k_A \to k_I = 1$. The superpotential has the form
\beq W = \sum_{IJK,ABC}c_{ABC}|\epsilon_{IJK}|\Phi^A_I\Phi^B_J\Phi^C_K.\eeq
This model has the property that 
\beq A_{IA,JB} = 0 \;\;{\rm if}\;\; I = J, \quad R^{i\m j\n}A_{ij} = 0, \eeq
where the indices $i,j,\cdots$ run over all chiral fields $z^i$, and 
the logarithmically divergent contributions (2.22-28) simplify
considerably.  However, the ansatz (\ref{skp}) is insufficient to cancel
logarithmic divergent terms proportional to $D_i(T^az)^jD_j(T_az)^i$ and
$D_i(T^az)^jR^j_{ik\m}$, suggesting that modular invariant
regularization is not possible for any choice of spectrum, although
invariance of the $O(\mu^2)$ term can always be imposed by conditions
analogous to (\ref{masses}). 

\subsection{String-derived supergravity}
If the underlying theory is a superstring theory, there is generally 
invariance under a discrete group of 
modular transformations on the light superfields under which $K\to K+ F(z) +
\bF(\z),\;\;W\to e^{-F(z)}W,$ which cannot be broken by perturbative quantum
corrections~\cite{mod}.  For example, in the class of orbifold
compactifications mentioned above the K\"ahler potential, including
twisted sector fields, takes the $SL(2,R)$ invariant form 
\bea K &=& \sum_{I=1}^3g^I + f\(e^\sum q^I_A|\Phi^A|^2\) =
\sum_{I=1}^3g^I + e^\sum q^I_A|\Phi^A|^2 + O\(|\Phi^A|^4\), \nonumber \\
g^I &=& - \ln\(T_I + \T_I\),\eea
which reduces to (\ref{noscale}) when the twisted fields are set to
zero. The general PV K\"ahler potential of (2.4) is modular invariant if
the field $Z^I_\alpha$ has the same modular 
weight as $Z^i$ and $\varphi^C$ has modular weight $\alpha_C$.
The superpotential (2.5) can be made invariant 
under the discrete $SL(2,Z)$ subgroup of $SL(2,R)$ modular transformations, 
by an appropriate $T_I$-dependence of the PV masses:
$\mu_\alpha \to \mu_\alpha(T_I) = \mu_\alpha\prod_I[\eta(T_I)]^{p^I_\alpha}$,  
where $\eta(T)$ is the Dedekind function. This modification of the
effective cut-offs could be interpreted as threshold effects arising from
the integration over heavy modes. 

On the other hand, it is known that at least some of the modular
invariance is restored by a universal Green-Schwarz counter
term; this is in particular the case for the anomalous Yang-Mills 
coupling~\cite{dixon}--\cite{tom}.  To study the conformal anomalies
arising from the noninvariance of the effective cut-offs, consider 
the helicity-even part\footnote{The chiral anomaly can be obtained by a
resummation~\cite{mksjr} of the derivative expansion of the helicity-odd
contribution $T_-$, which gives the standard results for the terms
condsidered here.}
of the one-loop action, given by
\beq S_1 = {i\over2}\STr\ln\[D^2 + H(M_{PV})\],\label{even}\eeq
where $M_{PV}$ is the PV mass matrix.  Under a transformation on the PV
fields, represented here by a column vector $X^i$, that leaves the
tree Lagrangian, as well as the PV K\"ahler potential, invariant:
\bea \pmatrix{\X^{\ibar}\cr X^i} \to g_i\pmatrix{\X^{\ibar}\cr X^i} ,\quad
M^{PV}_i = \pmatrix{0& m_i\cr \m_i&0\cr}, \quad 
M_{PV}\to M'_{PV} \nonumber\eea
\beq \(D^2 + H(0)\)_i \to g_i\(D^2 + H(0)\)_ig_i^{-1},\label{mi}\eeq
because all the operators in the determinant except $M_{PV}$ are covariant, and
the PV contribution to (\ref{even}) changes by
\bea \(S_1\)_{PV} &\to& {i\over2}\sum_i\eta_i
\STr\ln\lbr g_i\[D^2_i + H_i(g_i^{-1}M'_{PV}g_i)\]g_i^{-1}\rbr  \nonumber \\ 
&=& {i\over2}\sum_i\eta_i\STr\ln\[D_i^2 + H_i(g_i^{-1}M'_{PV}g_i)\], \eea
where $\eta_i$ is the signature, and the last equality holds if the integrals 
are finite.  The PV K\"ahler potential $K_{PV} = k_{i\m}X^iX^{\m}$ is invariant
provided $k_{i\m}\to g_i^{-1}k_{i\m}\g_m^{-1},\;k^{i\m}\to g_ik^{i\m}\g_m.$
If the PV mass is introduced via a superpotential term
$W\ni \mu_{ij}X^iX^j,\;\mu= $ constant, the PV mass is
\beq m_i^{\m} = e^{K/2}K^{j\m}\mu_{ij},\quad m'^{\m}_i = e^{(K'-K)/2}
\g_mK^{j\m}g_jK_{j\n}m_i^{\n}. \eeq
If the transformation is abelian: $g_i = e^{\phi_i}$, and the metric is 
diagonal:
$K_{i\m}\propto \delta_{i\m}$, we just get
\bea m'^{\m}_i&=&e^{(K'-K)/2 + \bph_m + \phi_i}m_i^{\m},\quad g_i =
\pmatrix{e^{\bph_i} &0\cr 0&e^{\phi_i}\cr}, \nonumber \\
g_i^{-1}M_ig_i&=&e^{(K'-K)/2}
\pmatrix{0 & e^{2\phi_i}m_i\cr e^{2\bph_i}\m_i &0\cr}, \eea
if, {\it e.g.}, $\mu_{ij}\propto \delta_{ij}$.

If, following section 2, we introduce regulators $X^A,X'_A$ for 
$\Phi^A$ with signature-weighted average modular weights $-q^A_I$, and
$X^a$ for the gauge fields with average weights $q^a_I = 1/3,$ and the
superpotential term
\beq W_{PV} = \sum_A \mu_AX^AX'_A + \sum_a \mu_aX^aX_a, \quad m_i = e^{K/2 -
\sum_I q_I^ig^I}\mu_i, \eeq
under a modular transformation we have 
\bea g_i &=& \pmatrix{e^{-\sum_Iq_I^i\bF^I}&0\cr 0&e^{-\sum_Iq_I^iF^I}\cr},
\quad m'_i = e^{\sum_I\(1 - 2q_I^i\){\rm Re}F^I}m_i,\nonumber \\
g_i^{-1}M'_ig_i &=& \pmatrix{0&e^{-2i\sum_Iq_I^i{\rm  Im}F^I}m'_i\cr
e^{2i\sum_Iq_I^i{\rm  Im}F^I}\m'_i&0\cr}, \eea
the contribution (\ref{ym}) shifts by
\bea &&-{1\over64\pi^2}\delta \lbr F_a^2\[3C_a\ln(|m^2_a|) - 
\sum_AC^A_a\ln(|m^2_A|)\]\rbr + \cdots  
\nonumber \\ && \quad = -{1\over32\pi^2}\sum_I {\rm Re}F^IF_a^2
\[C_a - \sum_AC^A_a\(1 -2q^I_A\)\] + \cdots,\label{modan} \eea
and the conformal anomaly matches the chiral anomaly arising from the
axial currents 
\beq A_\mu^\lambda = \Gamma_\mu =  {i\over4}\(\D_\mu z^iK_i - {\rm h.c.}\),
\quad \(A_\mu^\Phi\)^A_B = -\Gamma_\mu + {i\over2}\(\D_\mu z^i\Gamma^A_{Bi}
- {\rm h.c.}\),\eeq
for gauginos and charged chiral fermions, respectively.  The Casimirs and
modular weights satisfy the sum rules:
\beq C^a - \sum_A(1 - 2q^I_A)C^a_A = C_{E_8} - b^I_a.\eeq
For orbifolds such as $Z_3$ and $Z_7$ that contain no N=2 supersymmetric
twisted sector~\cite{ant}, $b^I_a = 0$, the anomaly (\ref{modan}) is
completely cancelled by a Green-Schwarz term.  For other models the
residual anomaly is cancelled by string-loop threshold
effects~\cite{dixon} that can be incorporated in the present formalism by
making the $\varphi^a$ masses moduli-dependent:
\beq\mu^\varphi_\alpha\to \prod_I[\eta(T_I)]^{b^I_a}\mu^\varphi_\alpha.\eeq
Note that since the masses are not modular invariant, additional
conditions, analogous to (\ref{masses}), must be imposed to make the
quadratically divergent terms anomaly free.  Possibilities for
cancelling the remaining modular anomalies will be studied elsewhere.

\section{Anomalous $U(1)$}
\hspace{0.8cm}\setcounter{equation}{0}

In this section we include an anomalous $U(1)_X$ gauge factor: $\Tr T_X,\Tr
T_X^3\ne 0$.
To regulate a nonanomalous gauge theory we introduced
heavy vector-like pairs of states with gauge invariant masses.
Explicitly, under a gauge
transformation $X^A\to g_A X^A,\; X'_A\to g_A^{-1}X_A,\; 
\X^{\A}\to g_A^{-1}\X^{\A},\;\X'_{\A}\to g_A\X_{\A},\; M' = gMg^{-1}$,
{\it i.e.}, the mass matrix (\ref{mi}) is covariant, and no anomaly is
introduced by the regularization procedure.

However, the quadratically divergent piece contains the term 
\beq 2x^{-1}\D_aD_i(T^az)^i = 2x^{-1}\D_a\(\Tr T^a + 
\Gamma^i_{ij}(T^az)^j\).\eeq 
If Tr$T_a\ne 0$, one cannot regulate the quadratic divergences\footnote{In 
the context of renormalizable theories one can use dimensional
regularization or reduction and the quadratic divergence never appears.}
without
introducing a mass term for PV states $X^i$ with the {\it same} $U(1)_X$ charge
$q^i$. As a consequence the effective cut-off is noninvariant, which
gives the conformal anomaly counterpart to the chiral anomaly.

Thus, in addition to the PV regulators introduced in section 2, we
introduce chiral fields $X^i$ with signatures $\eta_i$
that carry only $U(1)_X$ charge $q_i$:
\beq K\to K + k^i, \quad k^i = f^i(Z^j,\Z^{\m})|X^i|^2 + O|X^i|^4,
\quad W\to W + \mu^i(X^i)^2  . \eeq
Their contribution
to the chiral $U(1)_X$ anomaly vanishes; the explicit breaking through the
mass terms cancels their contribution to the true anomaly.  

We have been working with the covariant superspace formalism 
of~\cite{bggm}, in which the vector potential\footnote{$iA_\mu \to ia_m =
\l{\cal A}_m\r$ in the notation of~\cite{bggm}.} $A_\mu$ is introduced as
the lowest component of an anti-hermetian one-form superfield, and
matter superfields $\Phi$ are defined to be covariantly chiral:
\beq \Db\Phi = 0,\quad \chi^\alpha = \l\Da\Phi\r,\eeq
where the covariant derivative $\D_M$ contains the gauge connection
${\cal A}_M$, and $M$ is a coordinate index in superspace. Under a
gauge transformation:
\beq {\cal A}_M\to {\cal A}_M - g^{-1}D_Mg, 
\quad \Phi^A\to g^{q^A_X}\Phi^A,\quad g^{-1} = g^{\dag}\label{gauge}. \eeq
The chiral Yang-Mills superfield $W^\alpha$ is obtained as a component
of the two-form ${\cal F}_{MN}$, which is the Yang-Mills field strength in
superspace. The authors of~\cite{bggm} point out that one can introduce
the commonly used Yang-Mills superfield potential $V_X$ such that
\beq W_\alpha = - {1\over4}\(\DbDb-R\)\Dc V_X,\eeq
where $R$ is an element of the supervielbein and $\DbDb-R$ is the chiral
projection, 
but this field  does not appear in the construction of the action which
is invariant under an additional gauge transformation
\beq \vx \to \vx' = \vx + {1\over2}\(\Lambda + \bL\),\label{delv}\eeq
that is independent of (\ref{gauge}).
Since the gauge invariant superpotential is invariant under the complex
extension of the gauge group, there is no conflict between (\ref{gauge})
and holomorphicity of the superpotential.

However, the superpotential (4.2)
changes by a nonholomophic function under $U(1)_X$ if $X^i \to g^{q_i}X^i$.
Therefore holomorphicity requires $X^i \to e^{-q_i\Lambda}X^i,\;\Lambda$
holomorphic, under a $U(1)_X$ gauge transformation.  To preserve gauge
invariance of the K\"ahler potential, we take $X^i$ chiral in the ordinary
sense, that is, we define $\D_M X^i = D_MX^i$, where $D_M$ contains no
gauge connection, and modify the K\"ahler potential (4.2) to read
\beq K\to K + k^ie^{2q_i\vx}. \eeq
As shown in Appendix C, one obtains  the standard Lagrangian when this
expression is evaluated in the Wess-Zumino gauge.  
This choice is not justified unless the full theory is gauge
invariant. In fact, we are interested in the special case in which the
$U(1)_X$ anomaly satisfies the ``universality'' condition
\beq {1\over3}\Tr T_X^3 = \Tr(T_XT^2_a) = {1\over24}\Tr T_X = 8\pi^2\delta_X,
\label{gs}\eeq
and -- in string derived supergravity -- is cancelled by a Green-Schwarz
term~\cite{u1gs}. 
Thus provided this term is included and evaluated in the WS gauge,
there is no ambiguity.

Including the fields $X^i$ we get a quadratically divergent contribution:
\beq \STr H \ni 2g^2d_X\(\sum_A q_A^X + \sum_i\eta_iq_i\). \label{uquad}\eeq
where the first term is the light field contribution and $d_X =
\sum_AK_Aq_A^X\phi^A,\;\phi^A = \l\Phi^A\r.$ Finiteness requires
\beq \sum_i\eta_iq_i = - \Tr T_X = - 192\pi^2\delta_X, 
\quad \sum_i\eta_iq_im^2_i = 0.\eeq
Once all the infinities are cancelled one gets a finite contribution that grows
with $\mu^2$.  Setting $\mu_i = \beta_i\mu$, we get a contribution of
the form (2.17) with now 
\beq \delta K = \sum_i\eta_im_i^2\ln\beta_i, \quad m_i^2=\beta_i^2
e^KK_{i\ibar}^{-2}|\mu_i|^2.\eeq
Taking, for example, the modular invariant form 
\beq k^i = e^{K/2}|X^i|^2,\quad \delta K = 
{\mu^2\over32\pi^2}\sum_i\beta_i^2\ln\beta_ie^{-4q_i\vx},\eeq
the correction to the bosonic Lagrangian is [see (\ref{dlk}) and
Appendix C]
\bea \Delta\L &=& \sqrt{g}
{\mu^2\over32\pi^2}\l\[{1\over16}\Da\(\DbDb - R\)\Dc - \hV\]\delta K\r
\nonumber \\ &=&  \sqrt{g}{\mu^2\over32\pi^2}\sum_i\beta_i^2\ln\beta_i
\[2g^2q_i\(d_X - q_iA_\mu A^\mu\) - \hV\]. \label{umu}\eea
Note that a mass term is induced for the anomalous
$U(1)_X$ gauge boson $A_\mu$. Thus if the full quantum theory is not
anomalous we must impose
\beq \sum_i\eta_iq_i^2\beta_i^2\ln\beta_i = 0.\eeq

The logarithmically divergent contribution from $X^i$ contains a term
\beq \L_X \ni  - {1\over64\pi^2}\sum_i\eta_i\ln|m_i|^2 q_i^2F_X^2
+\cdots. \label{ulog}\eeq
Under $U(1)_X$, (\ref{delv}), $|m_i|^2 \to e^{-2q_i(\Lambda +
\bL)}|m_i|^2$, so the quantum Lagrangian changes by
\beq \delta\L_X \ni {1\over32\pi^2}\sum_i\eta_i\(\lambda + 
\lambda^*\)q_i^3F_X^2 +\cdots,\label{confx}\eeq
where $\lambda = \l\Lambda\r$.
The light fermion contribution gives the chiral anomaly:
\bea \delta\L_X &=& {i\dx\over2}\(\l\ln g\r\)\sum_aF^a{\tilde F_a} + 
\cdots \label{chix},\eea
For $F^a = F_X$, the anomalies (\ref{confx},\ref{chix}) form a supermultiplet 
if we take
\beq g = e^{-{i\over2}(\Lambda - \bar{\Lambda})}, \quad
\sum_i\eta_iq^3_i = 8\pi^2\dx.\eeq

To make the full anomaly determined by (\ref{gs}) supersymmetric, we
must include PV fields with both $U(1)_X$ and the nonanomalous gauge charges.
This can be accomplished by assigning the {\it same} $U(1)_X$ charge $q_A$ to 
the previously introduced PV fields $X^A,X'_A$, defining the superspace
derivative as $\D_M = D_MX + T_aA^a_M,\; A^a \ne A^X,$ and setting
$$|X^A|^2\to e^{2q_AV_X}|X^A|^2,\quad |X_A|^2\to e^{2q_AV_X}|X_A|^2,$$
in the K\"ahler
potential. The generalization of the Lagrangian of Appendix C to this
case is tedious but straightforward.  Once supersymmetry of the anomaly
is imposed, with the appropriate constraints on the PV $U(1)_X$ charges,
the full anomaly is cancelled by a Green-Schwarz term that gives the
variation of the Lagrangian under the $U(1)_X$ transformation (\ref{delv}):
\bea \delta\L^X_{GS} &=&-{\dx\over4}\int{E\over R}\Lambda\WaWa +
{\rm h.c.}
\nonumber \\ &=& -{\dx\over2}\({\rm Re}\lambda\sum_aF^aF_a +
{\rm Im}\lambda\sum_aF^a{\tilde F_a}\) + \cdots.\eea
This mechanism introduces a D-term with a well-defined coefficient that
has been used in many applications to phenomenology.  Note that there is
also a D-term in (\ref{umu}), that may be removed by an additional
condition on the $\beta_i$.  One needs further information on the
underlying theory to determine whether or not this term is present.

\section{Concluding remarks}
We have shown that on-shell one-loop Pauli-Villars regularization is
possible for supergravity theories with canonical kinetic energy for gauge
superfields.  The resulting Lagrangian depends on the PV masses $\mu$
that play the role of effective cut-offs.
It remains an open question as to whether PV regularization
remains possible at
higher order without the addition of higher derivative terms. However
since the chiral anomalies of the effective field theory are completely
determined at one loop order, and their partner conformal anomalies are
thereby fixed  by supersymmetry -- through constraints on the
Pauli-Villars massess -- at the same order, one loop calculations
are sufficient to study the field theory anomalies.  

We found
that nonlinear sigma-model symmetries can be preserved at the
quantum level only for ungauged theories with restricted particle spectra,
such that there are no quadratic divergences.  It is nevertheless
possible to impose invariance of the $O(\mu^2)$ correction, thereby
preserving the Bagger-Witten condition at the quantum level.  Similarly,
the $O(\mu^2)$ correction to an anomalous $U(1)$ gauge symmetry may be
made gauge invariant. There is also an $O(\mu^2)$ D-term that does not
automatically vanish when gauge invariance is imposed; further
information on the underlying theory is needed to fix this term.

In string-derived supergravity a discrete subgroup of the sigma-model
symmetry is preserved to all orders in perturbation theory; a study of
the anomaly structure provides information on the type of counterterms
that must be included to cancel the field theory anomalies.  In these
theories the gauge kinetic energy term is noncanonical, and is governed
by couplings to a universal dilaton.  The full loop corrections
including the dilaton, and a more detailed study of supergravity
theories based on orbifold compactifications of the heterotic string,
will be presented elsewhere.

\vskip .3in
\noindent{\bf Acknowledgements.}  I wish to thank Brent Nelson for his
help, Keith Deines for inspiration, and the LPTHE at Orsay, where this
work was completed, for hospitality.  This work was supported in part by the
Director, Office of Energy Research, Office of High Energy and Nuclear Physics,
Division of High Energy Physics of the U.S. Department of Energy under Contract
DE-AC03-76SF00098 and in part by the National Science Foundation under grant
PHY-95-14797. 

\vskip .3in
\appendix

\def\ksubsection{\Alph{subsection}}
\def\theequation{\ksubsection.\arabic{equation}} 

%       reset section commands
     
\catcode`\@=11

\def\thesubsection{\Alph{subsection}.}
\def\thesubsubsection{\arabic{subsubsection}.}
\noindent{\large \bf Appendix}

\subsection{The metric tensor for $Y$}
\setcounter{equation}{0} 
The metric tensor derived from $K_{\alpha>3}^Y$ in (2.4) is the inverse
of that derived from the K\"ahler potential 
\beq k =\[\sum_{I,J=i,j}Y_\alpha^{I}\Y_\alpha^{\J}\(K_{i\bj} 
+ a^2_\alpha K_iK_{\bj}\) + a_\alpha\(Y_\alpha^I\Y_\alpha^0K_i + {\rm
h.c.}\) + |Y_\alpha^0|^2 \] .\eeq
It is straightforward to evaluate the derivatives of the metric
$k_{P\bar{Q}}$, $P,Q=Y_I,Y_0$.
Denoting by $\gamma^P_{Qi},r^P_{Qi\m}$ the corresponding elements of the affine
connection and Riemann tensor, respectively, we have
\bea (T_a)^{Y_J}_{Y_I} &=& - (T_a)^i_j, \quad D_{Y_I}(T_aY)_J = -
D_j(T_az)^i, \nonumber \\
(\Gamma^Y)^Q_{Pi} &=& -\gamma^P_{Qi},\quad (R^Y)^Q_{Pi\m} = - r^P_{Qi\m}, \eea
giving the results listed in (2.13) and (2.42).   In addition we have,
\bea A^Y_{PQ} &=& e^KW_{PQ}, \quad \A_Y^{PQ} = 
e^KK_Y^{P\bar{P'}}K_Y^{P\bar{Q'}}\A_{\bar{P'}\bar{Q'}} =
e^Kk_{P\bar{P'}}k_{Q\bar{Q'}}\A_{\bar{P'}\bar{Q'}}, \nonumber \\
A^Y_{P\varphi} &=& e^KW_{P\varphi}, \quad \A^{P\varphi} = 
e^KK^{P\bar{Q}}K^{\varphi\bv}\A_{\bar{Q}\bv} = k_{P\bar{Q}}\A_{\bar{Q}\bv}, 
\eea
giving the results listed in (2.37).

\subsection{Nondiagonal gauge kinetic function}
\setcounter{equation}{0} 
Here we sketch the generalization of~\cite{us2} to the case of a nondiagonal
gauge kinetic function involving Pauli-Villars fields.  Although in this
paper, we assume a canonical kinetic energy term for the light gauge fields,
we give the results here for the case of a universal dilaton.  The case
relevant to section 2 of this paper is recovered by setting $s=$ constant.
With an arbitrary kinetic function $f_{ab}(Z)$, the Lagrangian for the
auxiliary fields $D_a$ of the Yang-Mills supermultiplets takes the
form~\cite{bggm}, upon solving for $D_a$,
\bea \L_D &=& {1\over2}\({\rm Re} f\)^{ab}D_aD_b - D_a\tD^a = - 
{1\over2}\[\({\rm Re} f\)^{-1}\]^{ab}\tD_a\tD_b , \nonumber \\
\tD^a &=& \D^a + {i\over2}\(f_i^{ab}\bar{\lambda}_b L\chi^i - {\rm h.c.}\),
\quad f_i^{ab} = \pp_if^{ab}. \eea
Writing $f^{ab} = f_a\delta^{ab} + \epsilon^{ab}$, we may expand in 
$\epsilon$ to obtain 
\beq \L_D = - {1\over2}\({\rm Re}f_a\)^{-1}\(\delta^{ab} - {{\rm Re}
\epsilon^{ab}\over{\rm Re}f_b} + 
\sum_c{{\rm Re}\epsilon^{ac}{\rm Re}\epsilon^{cb}\over{\rm Re}f_b{\rm Re}f_c}
\)\tD_a\tD_b + \cdots.\eeq
Here we introduce Pauli-Villars abelian gauge multiplets $W_\alpha^0$, and
take gauge kinetic functions of the form
\bea f^{AB} &=& \delta^{AB}\(x_A + iy_A\) + \epsilon^{AB}, \nonumber \\
f^{ab} &=& \delta^{ab}s + {d\over2}\varphi^a\varphi^b, \quad 
f^{\alpha\beta}_0 = \delta^{\alpha\beta}, \nonumber \\ 
f_\alpha^{a0} &=& e_\alpha\varphi^a, 
\quad K_{PV} = e^k\sum_a|\varphi^a|^2,\quad  e^k = {1\over2x} .\eea
In addition to scalar curvature terms, 
\bea R^a_{bs\s} = K_{s\s}\delta^a_b ,\eea
we have, for fixed $\alpha$,
\bea D_sf^{a0}_{\varphi^b} &=& -\Gamma^{\varphi^c}_{s\varphi^b}
f^{a0}_{\varphi^c} = - k_s \delta^a_be = {1\over2x}\delta^a_be .\eea

The relevant part of the tree Lagrangian~\cite{crem},~\cite{bggm} 
is (setting all background fermions to zero)
\bea 
{1\over\sqrt{g}}\L(\varphi^a,B_\mu^\alpha) &=& e^k\D^\mu\varphi^a\D_\mu\bv^a 
- {d\over16}\[\varphi^a\varphi^b\(F^b_{\mu\nu}F_a^{\mu\nu} 
- i{\tilde F}_a^{\mu\nu}F^b_{\mu\nu}\) + {\rm h.c.}\]\nonumber \\ & &
- {1\over4}F^\alpha_{\mu\nu}F_\alpha^{\mu\nu} 
 - {e_\alpha\over4}\[\varphi^a\(F^\alpha_{\mu\nu}F_a^{\mu\nu} 
- i{\tilde F}_a^{\mu\nu}F^\alpha_{\mu\nu}\) + {\rm h.c.}\] \nonumber \\ & & 
+ {i\over 2}\bl^\alpha\notD\lambda_\alpha + ie^k\(\bc_L^{a}\notD\chi_L^a +
\bc_R^a\notD\chi_R^a\) \nonumber \\ & &  - V
- e_\alpha\[i\bl^\alpha_R\({1\over2x}\D_a +
{1\over 4}\sigma_{\mu\nu}F^{\mu\nu}_a\)\chi^a_L + {\rm h.c.}\],\nonumber
 \eea \bea
V &=& - {1\over8x^2}\D_a\D_b\[d\(\varphi^a\varphi^b + \bv^a\bv^b\)
- e^2(\varphi^a + \bv^a)(\varphi^b+ \bv^b)\].\eea

Following the procedure described in~\cite{noncan}, we introduce
off-diagonal connections in the bosonic sector so as to cast the quantum
Lagrangian in the form
\bea \L_{\rm bose} + \L_{gh} &=& -{1\over 2}\Phi^T Z_\Phi\(\hD_\Phi^2 + H_\Phi\)
\Phi + {1\over 2}\bar{c}Z_{gh}\(\hD^2_{gh} + H_{gh}\)c, \nonumber \\
\hD_\mu^\Phi &=& D_\mu + V_\mu, \;\;\;\; \(V_\mu\)_{A\rho,B\sigma} = -
\epsilon_{\rho\mu\sigma\nu}{\pp^\nu y_{AB}\over2x},\nonumber \\
\(V_\mu\)_{A\nu,i} &=& \(V_\mu\)_{i,A\nu} = \[\(V_\mu\)_{\ibar,A\nu}\]^*=
{1\over4\sqrt{x_Ax_B}}f^{AB}_i\(\F_{B\mu\nu} - i\tcF_{B\mu\nu}\) \nonumber \\ 
&=& {e\over4}\(F_{b\mu\nu} - i\tF_{b\mu\nu}\)
, \quad {\rm for} \quad i = \varphi^b, \;\; A = A_\mu^0.\eea
This introduces corresponding shifts in the background field-dependent
``squared mass'' matrices:
\beq M^2_\Phi \to H_\Phi = M^2_\Phi - V_\mu V^\mu, \;\;\;\;
M^2_{gh} \to H_{gh} = M^2_{gh} - B_\mu B^\mu.\eeq

We have the following relations among derivatives of the kinetic function:
\bea f_a &=& D_af = e, \;\;\;\; f^a = 2xe, \quad
f_{sa} = D_sD_af = {e\over 2x}, \nonumber \\ f_{\s a} &=& D_{\s}f_a 
= 0, \quad R^a_{s\s b}f_aX^{s\s} = - {e\over4x^2}X^{s\s}, \nonumber \\
D_\mu e^2 &=& D_\mu\({f^{b\alpha}_{a}\f^{a}_{b\alpha}\over2x}\) 
\nonumber \\ &=& {1\over2x}\[\pp_\mu s\(D_sf^{b\alpha}_a\)\f^a_{b\alpha} 
+ {\rm h.c.}\] - {e^2\over2x^2}\pp_\mu x = 0, \label{rels}\eea

In evaluating the matrix elements needed for PV loop contributions, we set
background PV fields to zero and show explicitly only the terms involving the
parameters $e$ and $d$.  The remainder of this Appendix closely parallels
Appendix C of~\cite{us2}.

\subsubsection{Matrix elements}
The elements of $H_{IJ},\; I,J = \varphi^a$, are 
\bea H_{IJ} &=& \hV_{IJ} + R_{IJ} + \D_{IJ} + v_{IJ} - 
\(V_\mu V^\mu\)_{IJ}, \nonumber \\ 
v_{i\m} &=&  v_{\m i} = \(V_\mu V^\mu\)_{i\m} = \(V_\mu V^\mu\)_{\m i} = 0, 
\nonumber \\  \(V_\mu V^\mu\)_{\varphi^a\varphi^b} &=& {e^2\over8}
\(F^{\mu\nu}_aF^b_{\mu\nu}\mp i\tF^{\mu\nu}_aF^b_{\mu\nu}\),\nonumber \\
v_{\varphi^a\varphi^b} &=&  = {d\over8}\(F^{\mu\nu}_aF^b_{\mu\nu}\mp 
i\tF^{\mu\nu}_aF^b_{\mu\nu}\),\eea
where 
\bea \D_a^b &=& {e^2\over 2x}\D_a\D^b + {1\over x}\D_c(T^c)^b_a, \quad
\D_{ab} = {1\over 4x^2}\(e^2 - d\)\D_a\D_b.\eea
The additional nonvanishing elements of $Z_\Phi H_\Phi$
are $-N_{\alpha\mu,\beta\nu}$ and $S_{\alpha\mu,a}$ with
\bea N_{\alpha\mu,\beta\nu} &=& - {x\over2}e_\alpha e_\beta
\(F^a_{\mu\rho}F_{a\nu}^{\;\;\;\;\rho} - {1\over4}g_{\mu\nu}
F^a_{\rho\sigma}F_a^{\rho\sigma}\) + \delta_{\alpha\beta}r_{\mu\nu},
\nonumber \\ S_{\alpha\mu,a}^0 &=& {e_\alpha\over4}\D^\nu F_{a\nu\mu} +
{e_\alpha\over8x}F_{a\nu\mu}^{\mp}\pp^\nu\pmatrix{s\cr\s\cr}\nonumber \\
&=& {e_\alpha\over4}\[\hcD^\nu F_{a\nu\mu} -
{1\over2x}F_{a\nu\mu}^{\pm}\pp^\nu\pmatrix{\s\cr s\cr}\],\;\;\;\;a =
\cases{\varphi^a\cr\bv^a\cr},\nonumber \\
\hcD^\nu F_{a\nu\mu} &=& \D^\nu F_{a\nu\mu} + {\pp^\nu x\over x}F_{a\nu\mu} 
+ {\pp^\nu y\over x}\tF_{a\nu\mu} . \eea
Finally we need 
\bea \hG_{\mu\nu} &=& \(G_z + G_g + G_{gz}\)_{\mu\nu}, 
\nonumber \\ 
\(G^z_{\mu\nu}\)^a_b &=& {\pp_\mu\s\pp_\nu s - \pp_\mu\s\pp_\nu s\over4x^2}
\delta^a_b \pm iF^c_{\mu\nu}(T_c)^a_b,
\;\;\;\; a,b = \cases{\varphi^{a,b}\cr\bv^{a,b}\cr}, \nonumber \\ 
%\(G^2_{\mu\nu}\)^a_b &=& {\pp_\mu\s\pp_\nu s - \pp_\mu\s\pp_\nu s\over4x^2}
%\delta^a_b \pm iF^c_{\mu\nu}(T_c)^a_b, \;\;\;\; a,b = \cases{\varphi^{a,b}\cr
%\bv^{a,b}\cr}, \nonumber \\ 
\(G^z_{\mu\nu}\)_a^b &=& {e^2x\over4}\(F_{a\mu\rho}F_\nu^{b\;\rho} \mp 
i\tF_{a\mu\rho}F_\nu^{b\;\rho}\) - (\mu\leftrightarrow\nu),  
\;\;\;\; a,b = \cases{\varphi^a,\bv^b\cr\bv^a,\varphi^b\cr}, \nonumber \\ 
\(G^{gz}_{\mu\nu}\)_{\alpha\rho,a} &=& \(G^{gz}_{\mu\nu}\)_{a,\alpha\rho }
 \nonumber \\ &=& 
{e_\alpha\over4}\D_\mu F_{a\nu\rho}^{\mp} + {e_\alpha\over8x}
\pp_\mu\pmatrix{s\cr \s\cr}F_{a\nu\rho}^{\mp} 
-(\mu\leftrightarrow\nu),\;\;\;\;a = \cases{\varphi^a\cr\bv^a\cr},\nonumber \\ 
\(G^g_{\mu\nu}\)_{\alpha\rho,\beta\sigma} &=& 
\delta_{\alpha\beta}r_{\sigma\rho\mu\nu} + 
{x\over4}e_\alpha e_\beta\[F_{a\mu\rho}F^a_{\nu\sigma} + 
\tF_{a\mu\rho}\tF^a_{\nu\sigma} - (\mu\leftrightarrow\nu)\] . \eea 

The matrix elements of $M_\Theta$ are given by 
\bea M^0_0 &=& 0, \nonumber \\ 
M^0_a &=&  m_a + M^{\mu\nu}_a\sigma_{\mu\nu}, \;\;\;\; M_0^a 
= {1\over2}e^{-k}\(m_{a} - M^{\mu\nu}_{a}\sigma_{\mu\nu}\), 
\nonumber \\  m_{\varphi^a} &\equiv& m_a = {ie\over 2x}\D_a = m^*_{\bv^a} 
\equiv \m_a^* , \nonumber \\ 
M^{\mu\nu}_{ a} &=&  -{ie\over8}\(
F^{\mu\nu}_a \mp i\tF^{\mu\nu}_a\),\;\;\;\;a =\cases{\varphi^a\cr\bv^a\cr}, 
\eea 
with covariant derivatives as defined in~\cite{us,us2}:
\bea \D_\rho m_a &=& \D_\rho\(\m_a\)^* = -i{\pp_\rho\s\over4x^2}e\D_a
+ {ie\over2x}\(K_{j\m}(T_a\z)^{\m}\D_\rho z^j + {\rm h.c.}\), \nonumber \\ 
\D_\rho M^{\mu\nu}_a &=& -\(\D_\rho\bM^{\mu\nu}_a\)^* = -{ie\over8}
\(\D_\rho + {\pp_\rho s\over 2x}\)\(F_a^{\mu\nu} -i\tF_a^{\mu\nu}\) .\eea
The matrix elements of $G^\Theta_{\mu\nu}$ are (see Appendix D)
\bea \(G^{\pm}_{\mu\nu}\)_{00} &=& \pm\hGa_{\mu\nu} + Z_{\mu\nu}, 
\quad \hGa_{\mu\nu} = \Gamma_{\mu\nu} -{i\over2}F^a_{\mu\nu}\D_a,
\nonumber \\ \(G^\chi_{\mu\nu}\)^a_b &=& \(G^z_{\mu\nu}\)^a_b
+ \delta^a_b\(Z_{\mu\nu} \pm \hGa_{\mu\nu}\),
\;\;\;\;a,b=\cases{\varphi^a,\varphi^b\cr\bv^a,\bv^b}. \eea 

As in~\cite{us2}, we double the quantum fermions degrees of freedom and
represent them as 8-component Dirac spinors.  In the following $\btr$ denotes
the full trace of fermion mass and field strength ($G_{\mu\nu} =
[D_\mu,D_\nu]$) which are $8n_1\times 8n_2$ matrices, where $n_i$ is the
number of intrinsic fermion degrees of
freedom.  The explicit calculation given below is for just one nonvanishing 
$e^\alpha:\;\;n_i = N_G(1)$ for $\chi^a(\lambda^\alpha)$.

\subsubsection{Chiral multiplet supertrace}

Defining 
\beq {1\over 2}\STr H_\chi^2 = H^i_jH^j_i + H_{ij}H^{ij} - {1\over8}\btr
\(H_\Theta^{IJ}H^\Theta_{IJ}\),\;\;\;\; h^\chi_{\m i} = (\m m)_{\m i},\eeq
we have 
\bea {1\over8}\btr\(H_1^\chi\)^2 &=& \Tr \;h_\chi^2 + {e^4\over32}
\D_a\D^bF^a_{\mu\nu}F_b^{\mu\nu}, 
\quad (h^\chi)_a^b = {e^2\over4x}\D_a\D^b , \nonumber \\
H_a^b &=& (h^\chi)_a^b + \delta_a^b\(\hV + M^2 - M^2_\lambda 
- {\pp_\mu s\pp^\mu\s\over4x^2}\) 
+ {e^2\over4x}\D_a\D^b + {1\over x}\D_c(T^c)^b_a, \nonumber \\ 
H_{ab} &=& {1\over2}\(d - e^2\)\cW_{ab}. \eea 
Thus: 
\bea {1\over8}\btr\(H_1^\chi\)^2 &=& \Tr \;h_\chi^2 + {e^4\over16}
\[\(\cW^{ab} + \cbW^{ab}\)\D_a\D_b + 4\D^2\], 
\nonumber \\ \btr\(H_2^\chi\)^2 &=& 0, 
\;\;\;\; {1\over8}\btr H_3^\chi = O(N_G), \nonumber \\
{1\over8}\btr\(H_3^\chi\)^2 &=& {1\over2}\Tr(T_aT^b)F^a_{\mu\nu}F_b^{\mu\nu}  
+ O(N_G), \nonumber \\ 
{1\over4}\btr H_3^\chi H_1^\chi &=& - T_3^\chi + {r\over2} \Tr \;h^\chi , \eea
where
\bea \Tr \;h^\chi &=& {e^2\over2}\D, \nonumber \\
T_3^\chi &=& {ie^2\over4}\D_\mu z^j\D_\nu\z^{\m}K_{i\m}\D^aF_a^{\mu\nu}
+ {ie^2\over4x^2}\D^aF_a^{\mu\nu}\pp_\mu s\pp_\nu\s \nonumber \\ & &
+ {e^2\over4}\D_a\D_b\(\cW^{ab} + \cbW^{ab}\) + e^2\D^2,\eea
and the chiral fermion contributions to the helicity-odd operator $T_-$
(see~\cite{us2}) are
\bea T^\chi &=& T^\chi_3 + T^\chi_4 + T^\chi_r, \nonumber \\
T^\chi_r &=& - {e^2\over12}x
\(r^\mu_\nu F_a^{\nu\rho}F^a_{\mu\rho} - {1\over4}rF_a^{\mu\nu}F^a_{\mu\nu}\), 
\nonumber \\ T_4^\chi &=& {e^4x^2\over384}
\[(F^a_{\mu\nu}F_b^{\mu\nu})^2 + (F^a_{\mu\nu}\tF_b^{\mu\nu})^2\] -
{e^4\over32}\D_a\D^bF^a_{\mu\nu}F_b^{\mu\nu}. \eea
Then we obtain \bea
\STr H_\chi &=& e^2\D + O(N_G), \nonumber \\ 
{1\over 2}\STr H_\chi^2 &=& - T_3^\chi - 
\(\cW_{ab} + \cbW_{ab}\)\[\Tr(T^bT^a) + {e^4\over16}\D_a\D_b\]  
- {e^2\over4}r\D \nonumber \\ & & 
+ {e^4\over2}\D^2 + 2e^2\(\hV + M^2 - M^2_\lambda\)\D 
\nonumber \\ & & - {e^2\over2x^2}\D\pp^\mu\s\pp_\mu s 
+ (d-e^2)^2x^2\cW_{ab}\cbW^{ab} \nonumber \\ & &
+ {ie^2\over2}\({\pp_\mu s\pp_\nu\s\over x^2} + 
\D_\mu z^j\D_\nu\z^{\m}K_{j\m}\)\D^aF_a^{\mu\nu} \nonumber \\ & & 
+ {e^2\over2}\D_a\D_b\(\cW^{ab} + \cbW^{ab}\) + 2e^2\D^2 + O(N_G). \eea
Finally we have
\beq
\(G^z_{\mu\nu}\)_{\varphi^a}^{\bv^b}\(G^z_{\mu\nu}\)_{\bv^b}^{\varphi^a}
= 0, \eeq so
\bea {1\over12}\STr\hG_{\mu\nu}^\chi\hG^{\mu\nu}_\chi &=& - T^\chi_r +
{e^2x\over24}\(
r^\mu_\nu F^a_{\mu\rho} F_a^{\nu\rho} - {r\over4}F^a_{\mu\nu} F_a^{\mu\nu}\) 
\nonumber \\ & & + {e^4\over384}\[\(F^a_{\mu\nu} F_b^{\mu\nu}\)^2 + 
\(F^a_{\mu\nu}\tF_b^{\mu\nu}\)^2\]+ O(N_G). \eea

\subsubsection
{Mixed chiral-gauge supertrace}
For the bose sector we have $H_\Phi^{\chi g} = -S$, and, using (B.17)
of~\cite{us2},
\bea \Tr S^2 &=& {e^2x\over4}\(\hcD_\nu F_a^{\mu\nu}\)^2 
+ {e^2\over16x}F_a^{+\nu\mu}F^{-a}_{\rho\mu}\pp^\rho s\pp_\nu\s 
- {e^2\over8}\(F^{-a}_{\nu\mu}\pp^\nu s + {\rm h.c.}\)\hcD_\rho F_a^{\rho\mu}  
\nonumber \\ &=& {e^2\over4x}\[g^{-{1\over2}}\L_a^\mu - 
iK_{i\m}\(\D^\mu\z^{\m}(T_az)^i - \D^\mu z^i(T_a\z)^{\m}\) \]^2 
+ {e^2\over16x}F_a^{+\nu\mu}F^{-a}_{\rho\mu}\pp^\rho s\pp_\nu\s \nonumber \\ & &
- {e^2\over8x}\(F^{-a}_{\nu\mu}\pp^\nu s + {\rm h.c.}\)
\[g^{-{1\over2}}\L_a^\mu - i\(K_{i\m}\D^\mu\z^{\m}(T_az)^i - {\rm h.c.}\)
\].   \eea
To evaluate the fermion matrix elements we use Eqs. (3.36) and (C.24-27)
of~\cite{us2}:
\bea {1\over 8}\btr\(H_1^{\chi g}\)^2 &=& 0, \nonumber \\
-{1\over 8}\btr\(H^{\chi g}_2\)^2 &=& 2\(\D_\mu\m\)^a_0\(\D^\mu m\)^0_a 
- 8\(\D_\mu\bM^{\mu\nu}\)^a_0\(\D^\rho M_{\rho\nu}\)^0_a ,\eea  
with 
\bea & & 8\(\D_\mu\bM^{\mu\nu}\)^a_0\(\D^\rho M_{\rho\nu}\)^0_a = 
{1\over2}\Tr S^2,\nonumber \\
& & 2\(\D_\mu\m\)^a_0\(\D^\mu m\)^0_a = {e^2\over4x^2}
\(\pp_\mu x\pp^\mu x + \pp_\mu y\pp^\mu y\)\D \nonumber \\ & & \qquad
+ {e^2\over2x} \lbr K_{i\n}K_{j\m}\D^\mu z^j
(T_a\z)^{\m}\[(T^az)^i\D_\mu\z^{\n} + (T^a\z)^{\n}\D_\mu z^i\] + {\rm h.c.}\rbr
\nonumber \\ & & \qquad
- {e^2\over2x^2} \pp^\mu x \D^a K_{j\m}\[
(T_az)^j\D_\mu \z^{\m} + (T_a\z)^{\m}\D_\mu z^j\],\eea
and
\beq T^{\chi g} = t_3^{\chi g} = - 
{16\over3}\(\D^\sigma\bM_{\sigma\mu}\)^a_0\(\D_\rho 
M^{\rho\mu}\)^0_a = - {1\over3}\Tr S^2.\eeq
In addition we have
\bea \Tr\(\hG_\Phi^{\chi g}\)^2 &=& 4\(G^{gz}_{\mu\nu}\)_{0\rho,a}
\(G^{gz}_{\mu\nu}\)^{a,0\rho } = {1\over2}\btr\(\hG_\Theta^{\chi g}\)^2 
\nonumber \\ &=& 64\(\D_\mu\bM_{\nu\rho}\)_0^a\(\D^\mu M^{\nu\rho} - 
\D^\nu M^{\mu\rho}\)_a^0 = - 4\Tr S^2.\eea
Using the classical equations of motion (B.17) of~\cite{us2}, we obtain, 
\bea L_{\chi g} &=& {1\over2}\STr H^2_{\chi g} + T_{\chi g} 
+ {1\over12}\STr\hG_{\chi g}^2 \nonumber \\ &=&
- {2e^2\over\sqrt{g}}\Delta_{\D}\L - {e^2\over 2gx}
\L_{a\mu}\L^{a\mu} + {e^2\over2x\sqrt{g}}\(F^a_{\nu\mu}\pp^\nu x + 
\tF^a_{\nu\mu}\pp^\nu y\)\L_a^\mu \nonumber \\ & & 
+ {e^2\over x\sqrt{g}}\[i\L^{a\mu}\(K_{i\m}\D_\mu\z^{\m}(T_az)^i - 
{\rm h.c.}\) + \D^a(T_az)^I\L_I\] \nonumber \\ & & + 2e^2\D\(2M^2 +
2M^2_\lambda + 2\re(M\bM_\lambda) + \hV \) \nonumber \\ & & + 
{5e^2\over4x^2}\D\pp_\mu s\pp^\mu\s 
- {e^2\over8x}\(F^{\nu\mu} - i\tF^{\nu\mu}\)\(F_{\rho\mu} + 
i\tF_{\rho\mu}\)\pp_\nu s\pp^\rho\s \nonumber \\ & &
- {e^2\over2x^2} \[\(\pp^\mu x + 2i\pp^\mu y\) 
K_{j\m}(T_az)^j\D_\mu \z^{\m} + {\rm h.c.}\]\D^a \nonumber \\ & & 
- {ie^2\over2x}\(F^a_{\nu\mu}\pp^\nu x + 
\tF^a_{\nu\mu}\pp^\nu y\)\[K_{j\m}(T_az)^j\D_\mu \z^{\m} - {\rm h.c.}\]
\nonumber \\ &=& - {2e^2\over\sqrt{g}}\Delta_{\D}\L - {e^2\over 2gx}
\L_{a\mu}\L^{a\mu} + {e^2\over2x\sqrt{g}}\(F^a_{\nu\mu}\pp^\nu x + 
\tF^a_{\nu\mu}\pp^\nu y - {\pp^\mu y\over x}\D^a\)\L_a^\mu \nonumber \\ & & 
+ {e^2\over x\sqrt{g}}\[i\L^{a\mu}\(K_{i\m}\D_\mu\z^{\m}(T_az)^i - 
{\rm h.c.}\) + \D^a(T_az)^I\L_I\] \nonumber \\ & & + 2e^2\D\(2M^2 +
2M^2_\lambda + 2\re(M\bM_\lambda)  + \hV \) 
+ {e^2\over x^2}\pp^\mu x\pp^\nu yF^a_{\mu\nu}\D^a\nonumber \\ & & + 
{5e^2\over4x^2}\D\pp_\mu s\pp^\mu\s 
- {e^2\over8x}\(F^{\nu\mu} - i\tF^{\nu\mu}\)\(F_{\rho\mu} + 
i\tF_{\rho\mu}\)\pp_\nu s\pp^\rho\s \nonumber \\ & &
- {e^2\over2x^2}\[\(ixF^{-a}_{\nu\mu}\pp^\nu s + \pp_\mu s \D^a \)
K_{j\m}(T_az)^j\D^\mu \z^{\m} + {\rm h.c.}\],\eea
where in the last expression (C.76) of~\cite{us2} was used with (\ref{rels}) 
above.

\subsubsection{Yang-Mills supertrace}

For the bosonic contributions, we have $H^g_\Phi = - N$;
we write $N_{\alpha\beta} = N'_{\alpha\beta} + \delta_{\alpha\beta}n$, 
and evaluate here only $N'_{\alpha\beta} \to N_{00}$.  
\bea \Tr N &=& N'_Gn, \nonumber \\
\Tr N^2 &=& N'_Gn^2 - e^2x\(r_\nu^\mu F^a_{\mu\rho}F_a^{\nu\rho} 
- {1\over 4}rF^a_{\mu\nu}F_a^{\mu\nu}\) \nonumber \\ & &
+ {x^2e^4\over16}\[\(F^a_{\mu\nu}F_b^{\mu\nu}\)^2
+ \(F^a_{\mu\nu}\tF_b^{\mu\nu}\)^2\] ,\eea
and, writing $\(\hG^g_{\mu\nu}\)^\alpha_\beta = 
\(\hG'_{\mu\nu}\)^\alpha_\beta + \hgg_{\mu\nu}\delta^\alpha_\beta$, 
\bea \Tr\(\hG^g_\Phi\)^2 &=& N'_G\hgg^2 
+ {xe^2\over2}\(4r_\mu^\nu F_a^{\mu\rho}F^a_{\nu\rho} -r_\nu^\mu
F^a_{\mu\rho}F_a^{\nu\rho} \) \nonumber \\ & &
- {x^2e^4\over8}\[\(F^a_{\mu\nu}F_b^{\mu\nu}\)^2
+ \(F^a_{\mu\nu}\tF_b^{\mu\nu}\)^2\] ,\eea
where we dropped total derivatives and used (B.12--B.14) of~\cite{us2}, as 
well as the Yang-Mills Bianchi identity.  For the fermions we obtain:
\bea {1\over8}\btr H^g_1 &=& {N'_G\over4}\btr h_1 + {e^2\over2}\D, 
\nonumber \\
{1\over8}\btr\(H^g_1\)^2 &=& {e^4\over4}\D^2 
+ {e^4\over32}F^a_{\mu\nu}F_b^{\mu\nu}\D_a\D^b, \nonumber \\
- {1\over8}\btr\(H^g_2\)^2 &=& 0, \quad 
{1\over8}\btr H^g_3 = {N'_G\over4}r, \quad 
{1\over8}\btr\(H^g_3\)^2 = {N'_G\over4}\btr h^2_3, \nonumber \\
{1\over4}\btr\(H_1H_3\)^g &=& {N'_G\over2}\Tr\(h_1h_3\)^g +
{e^2\over4}r\D - {ie^2\over4}\D_\mu z^j\D_\nu\z^{\m}K_{i\m}\D^aF_a^{\mu\nu} 
\nonumber \\ & &
- {e^2\over4}\D_a\D_b\(\cW^{ab} + \cbW^{ab}\) - e^2\D^2, \nonumber \\
{1\over2}\btr\hG^2_g &=& N'_G\btr\hgg^2
+ xe^2\(r^\mu_\nu F^a_{\mu\rho} F_a^{\nu\rho} - 
{r\over4}F^a_{\mu\nu} F_a^{\mu\nu}\) \nonumber \\ & & 
- {x^2e^4\over16}\[\(F^a_{\mu\nu} F_b^{\mu\nu}\)^2 
+ \(F^a_{\mu\nu}\tF_b^{\mu\nu}\)^2\]. \eea
The nonvanishing contributions to $T^g = T_3^g + T_4^g + T_r^g$ are:
\bea T_3^g &=& {ie^2\over4}\D_\mu z^j\D_\nu\z^{\m}K_{i\m}\D^a
F_a^{\mu\nu}  \nonumber \\ & &
+ {e^2\over4}\D_a\D_b\(\cW^{ab} + \cbW^{ab}\) + e^2\D^2 + N'_Gt_3, 
\nonumber \\ T^g_4 &=& T^\chi_4, \quad T_r^g = T_r^\chi. \eea
For the supertraces we obtain [using (B.17--20) of~\cite{us2}] 
\bea \STr H^g &=& N'_G\STr h^g - e^2\D,
\nonumber \\
{1\over2}\STr H^2_g &=& {1\over2}N'_G\STr h^2_g 
+ {e^4\over2}\[x^2\cW^{ab}\cbW_{ab} + {3\over8}\D_a\D_b
\(\cW^{ab} + \cbW^{ab}\)\] \nonumber \\ & &
- T^g_3 - {xe^2\over2}\(r_\nu^\mu F^a_{\mu\rho}F_a^{\nu\rho} 
- {1\over 4}rF^a_{\mu\nu}F_a^{\mu\nu}\) - {e^2\over4}r\D \nonumber \\ & & 
+ {ie^2\over2}K_{j\m}\D_\mu z^j\D_\nu\z^{\m}\D^aF_a^{\mu\nu}
+ {e^2\over2}\D_a\D_b\(\cW^{ab} + \cbW^{ab}\) + 2e^2\D^2, \nonumber \\ 
{1\over12}\STr\hG_g^2 &=& {1\over12}N'_G\STr\hgg^2 
- {1\over12}\STr\hG_{\chi}^2 - T_4^g - T_4^\chi - T_r^g - T_r^\chi
\nonumber \\ & & - {e^4\over8}\[4\D^2 + \D_a\D_b\(\cW^{ab} + \cbW^{ab}\)\].\eea

The space-time curvature dependent terms in the supertraces evaluated above
give a contribution $\L_r$ of the form (2.23) of~\cite{us} with
\bea H_{\mu\nu} &=& H^g_{\mu\nu}
- \lln e^2x \(F^a_{\mu\rho}F_{a\nu}^{\;\;\;\;\rho} - {1\over4}
g_{\mu\nu}xF^a_{\rho\sigma} F_a^{\rho\sigma} \), 
\nonumber \\ \epsilon_0 &=& \epsilon^g_0 -\lln e^2\D , \nonumber \\
\alpha &=& \alpha^g, \;\;\;\; \beta = \beta^g. \eea 
The metric redefinition in (2.24--25) of~\cite{us} gives a correction
\bea \Delta_r\L &=& \lln\Delta_r L, \nonumber \\ 
\Delta_r L &=& \Delta_{rg}\L + e^2\(\D_\mu z^i\D^\mu\z^{\m}K_{i\m} -2V\)\D 
\nonumber \\ & & 
+ e^2x\(F^a_{\rho\mu}F_a^{\rho\nu}\D_\nu z^i\D^\mu\z^{\m}K_{i\m} 
- {1\over4}F^a_{\rho\sigma}F_a^{\rho\sigma}\D_\mu z^i\D^\mu\z^{\m}K_{i\m}\)
\nonumber \\ & & 
- 2e^2\[x^2\cW^{ab}\cbW_{ab} + {1\over2}\(\cW^{ab} + 
\cbW^{ab}\)\D_a\D_b + \D^2\] .\eea 

The result for constant $x$, given in (\ref{lw}) of section 2, is obtained by 
setting $M_\lambda = 0,\;y=0,\;s=x=g^{-2}$ constant in the above equations.  
In section 2 the fields $\hph_\alpha^a$ are taken to be canonically 
normalized. Combining the above results and 
evaluating $\L_1 - \L_r + \Delta_r\L - \Delta_K\L -
\Delta_x\L - \L_AX^A - \L_A\L_BX^{AB}$ yields the results given in 
(\ref{lw}), with $\varphi^a\to\sqrt{2x}\hph^a_\alpha$  and
\bea e^2 &\to& \sum_{\beta\gamma}\eta^{\hph}_\gamma e^2_{\gamma\beta}\equiv 
2e,\quad e^4 \to \sum_{\alpha\beta\gamma\delta}\eta^{\hph}_\gamma
e_\alpha^\beta e_\beta^\gamma e_\gamma^\delta e^\alpha_\delta\equiv 4e^2,
\nonumber \\ (d-e)^2 &\to& \sum_{\beta\gamma}\eta^{\hph}_\gamma\(d_{\gamma\beta} 
- \sum_\alpha e_{\gamma\alpha}e_{\alpha\beta}\)^2\equiv (d - 2e)^2.\eea

\subsection{Lagrangian with a vector potential superfield}
\setcounter{equation}{0} 
In this appendix we follow the notation of~\cite{bggm}: Greek letters
are used for two-component spinorial indices, Roman letters for tangent
space and coordinate indices, and the metric is $(-+++)$, {\it i.e.} the
negative of the one used elsewhere in the text.  We include the chiral fields
$X^x = \{X^i,Z^a\}$, where the $X^i$ are PV regulator fields charged only under
an anomalous $U(1)_X$, and $Z^a$ are the physical, light fields of the
effective low energy theory.
 
Defining, in analogy with the chiral superfield 
$X_\alpha = -{1\over8}\(\DbDb - 8R\)\D_\alpha K$ introduced in~\cite{bggm},
\beq X'_\alpha = -{1\over8}\(\DbDb - 8R\)\D_\alpha\(k^ie^{2q_i\vx}\), \quad
x_\alpha = -{1\over8}\(\DbDb - 8R\)\D_\alpha k^i, \eeq
the PV Lagrangian gets contributions (in WZ gauge)
\bea \L^i_{PV}&\ni&- \l{1\over4}\D^\alpha X'_\alpha\r + \l{i\over2}\bar{\psi}_m
\bar{\sigma}^m X'\r + {\rm h.c.} = - \l{1\over4}\D^\alpha x_\alpha\r + 
\l{i\over2}\bar{\psi}_m \bar{\sigma}^m x\r \nonumber \\ & &
+ {1\over2}k^iq_i\bar{\psi}_m\bar{\sigma}^m\lambda_X 
- i{\sqrt{2}\over2}\bar{\psi}_m \bar{\sigma}^n\sigma^m\bc^{\x}k^i_{\x}a_m
-{1\over2}q_i^2k^ia_ma^m + iq_ia_m\D^mz^xk^i_x \nonumber \\ & & 
+ {1\over2}q_ik^iD_X + q_i\sqrt{2}\chi^x\lambda_Xk^i_x +
{1\over2}q_ia_mk^i_{x\y}\bc^{\y}\sigma^m\chi^x 
+ {\rm h.c.} \nonumber \\ &=&- \l{1\over4}\D^\alpha x_\alpha\r + 
\l{i\over2}\bar{\psi}_m \bar{\sigma}^m x\r 
+ {1\over2}d^i\bar{\psi}_m\bar{\sigma}^m\lambda_X 
- i{\sqrt{2}\over2}\bar{\psi}_m \bar{\sigma}^n\sigma^m\bc^{\x}K'_{i\x}
(T_Xx)^ia_m \nonumber \\ & &
-{1\over2}(T_Xx)^i(T_X\x)^{\ibar}K'_{i\ibar}a_ma^m 
+ 
ia_m\D^mz^x(T_X\x)^{\ibar}K'_{x\ibar} +{1\over2}d^iD_X \nonumber \\ & &
+ \sqrt{2}\chi^x\lambda_X(T_X\x)^{\ibar}K'_{x\ibar} +
{1\over2}q_ia_mk^i_{x\y}\bc^{\y}\sigma^m\chi^x + 
{\rm h.c.} ,\eea 
where $K' = K + k^i$ and the last equality follows because 
\beq q_ik^i = (T_Xx)^iK'_i = d^i, \quad q_ik^i_{\x} = (T_Xz)^iK'_{i\x},
\quad q^2_ik^i = (T_Xx)^i(T_X\x)^{\ibar}K'_{i\ibar} . \eeq
The first two terms are the contributions to ${\tilde\D}_{\cal M}$
of~\cite{bggm} quadratic in $X^i$ without the gauge connections for $X^i$, and
\beq k^i_i = {\pp k^i\over x^i}, \quad k^i_a = {\pp k^i\over z^a}, \quad etc.,
\quad x^i=\l X^i\r,\quad a\ne i, \quad x,y = i,a.\eeq
The remaining terms covariantize $\pp_m x^i$ and give the correct
$\psi,\lambda_X, 
D_X$ terms. All fermion derivatives include the K\"ahler $U(1)$ connection that
has a piece:
\bea \l A_a\r &\ni& {1\over16}\sba\l \[\Dc,\Db\]\(k^ie^{2q_i\vx}\)\r
= {1\over16}\sba\l \[\Dc,\Db\]k^i\r + {i\over2}q_ik^iv_a \nonumber \\
&\ni& {1\over4}K'_i\(\pp_a + iq_ia_a\)x^i - {\rm h.c.}.\eea
In other words $A_a$ includes the gauge connection for $U(1)_X$.
The fully covariant derivative for the fermions contains the additional gauge
connection terms:
\bea D_m \chi^x&\ni&ia_m\[\(T_X\chi\)^x + \(T_Xz\)^y\Gamma'^x_{yz}\chi^z\], 
\nonumber \\
D_m \chi^i_X&\ni&ia_mq_i\(\chi^i_X + x^i\Gamma'^i_{ia}\chi^a\) + O(X^3)
= ia_mq_i\(\chi^i_X + {k^i_{a\ibar}\over k_{i\ibar}^i} \chi^a\) + O(X^3) 
\quad\quad\nonumber \\ 
D_m \chi^a&\ni&ia_mq_ix^i\Gamma'^a_{ib}\chi^b + O(X^4) = ia_mq_iK^{a\bb}
\(k^i_{\bb b}- {k^i_{\bb i}k^i_{b\ibar}\over k^i_{\ibar i}}\)\chi^b + O(X^4), 
\quad\quad \eea 
where we used the fact that
\beq K'^{a\ibar} = - K^{a\bb}{k^i_{\bb i}\over k^i_{\ibar i}} + O(X^3),\eeq
So the fully covariant kinetic energy term contains the terms: 
\beq - {i\over 2}\(D_m\chi^x\)\chi^{\y}K'_{x\y} + {\rm h.c.} \ni q_ia_m
k^i_{x\y}\bc^{\y}\sigma^m\chi^x + {\rm h.c.} + O(X^4) ,\eeq
which is just the last term in (C.2). Thus we get the
standard form of the tree Lagrangian, and loop corrections from $X^i$
are also of 
standard form.  Converting to the notation used previously ({\it e.g.},
$a_ma^m\to -A_\mu A^\mu),$ we obtain the results
(\ref{uquad},\ref{umu},\ref{ulog}) given in section 4, where we used the
classical equation of motion $D_X = - g^2d_X.$  The right hand side of 
(\ref{umu}) is given by the RHS of (C.2) with fermion fields set to
zero and $k^i\to \mu^2=$ constant.

\subsection{Errata}
\setcounter{equation}{0} 
Here we list corrections to~\cite{us,us2}. {\small{
\begin{enumerate}
\item The term $+{1\over 8}\(g_{\mu\rho}r_{\nu\sigma} + g_{\nu\rho}r_{\mu\sigma}
+ g_{\mu\sigma}r_{\nu\rho} + g_{\nu\sigma}r_{\mu\rho}\)$ is missing from the 
expression for $X_{\mu\nu,\rho\sigma}$ in (2.22) and (B3) of~\cite{us}.  As a
consequence (B6) should read
$$\Tr X = -20V + 2r, \quad \Tr X^2 = 40V^2 - 24rV + 22r_{\mu\nu}r^{\mu\nu} 
- 2r^2 + {\rm total\; derivative},$$
the following replacements should be made in (B20):
$$ {N + 1\over12}r^2\to {N - 7\over12}r^2, \quad - 5Vr\to - 13Vr, \quad
r_{\mu\nu}r^{\mu\nu}\to 8r_{\mu\nu}r^{\mu\nu}, $$
the first three equations in (B22) should read:
\bea \alpha &=& - 2\lln, \;\;\;\; \beta = {N+89\over 6}\lln,\nonumber\\
\epsilon_0 &=& -\lln\lbr e^{-K}\(A_{ij}\A^{ij} - {2\over 3}R^i_jA_i\A^j\)
 + {2N+68\over 3}\hV + {2N+16\over 3}M^2_\psi\rbr,\nonumber \eea
and (B23) (as well as footnote 23 of~\cite{us2}) should read: 
$$ {1\over\sqrt{g}}\Delta_r\L
 =  \lln\Bigg[\lbr -2e^{-K}\(A_{ki}\A^{ik} - {2\over3}R^k_nA_k\A^n\) 
- {3N +95\over3}\hV - {4N+32\over3}M^2_\psi\rbr\hV $$
$$ + \[K_{i\m}\lbr {N+55\over 3}\hV + e^{-K}\(A_{ki}\A^{ik} - 
{2\over 3}R^k_nA_k\A^n\) + {2N+16\over 3}M^2_\psi \rbr 
+{4\over 3}R_{i\m}\hV\]\D_\rho z^i\D^\rho\z^{\m}$$ $$
 - \lbr{2\over 3}\(R_{i\m} + 16K_{i\m}\)\D_\rho z^i\D^\rho\z^{\m}g_{\mu\nu} 
- {N+113\over 6}\(\D_\mu z^i\D_\nu\z^{\m} + \D_\nu z^i\D_\mu\z^{\m}\)K_{i\m}
\rbr\D^\mu z^j\D^\mu\z^{\n}K_{i\n} \Bigg].$$
In addition, in Eq. (C.55) of~\cite{us2}, the replacements
$$ xF_{\mu\nu}^aF_a^{\mu\nu}r \to 5xF_{\mu\nu}^aF_a^{\mu\nu}r, \quad
+ 2r^\mu_\nu xF_{\mu\rho}^aF_a^{\nu\rho}\to 
-12r^\mu_\nu xF_{\mu\rho}^aF_a^{\nu\rho} , $$
should be made in the expression for Tr$X^2$, and the replacements
$$ -{3x\over4}rF_{\mu\nu}^aF_a^{\mu\nu} \to +{5x\over4}r
F_{\mu\nu}^aF_a^{\mu\nu}, \quad
+ 2r^\mu_\nu xF_{\mu\rho}^aF_a^{\nu\rho}\to 
-5r^\mu_\nu xF_{\mu\rho}^aF_a^{\nu\rho} , \quad -5r\D\to -13r\D,$$
should be made in the second equation of (C.62).

\item In (3.33) the expression for $T_3$ is missing a term:
$$ T_3\to T_3 - {i\over3p^2}r^\mu_\nu\Tr\({\tilde M}_{\mu\rho}\bM^{\nu\rho} 
- {\tilde{\bM}}_{\mu\rho}M^{\nu\rho}\),$$ 
the last line of $\Tr {\cal R}{\cal R}_5$ in (3.35) has the wrong sign, and 
the last term in the second line of the RHS of (3.36) should be multiplied by  
$-2/3$.  As a 
consequence, ${1\over 8}\to -{1\over12}$ in $T_3^\chi$, (C.35), and in $T^g_3$,
(C.59); ${7\over 8}\to {13\over12}$ in the fourth line of (C.62). In
addition ${1\over 4}\to {1\over 6}$ in the second line of $\STr
\hat{G}^2_{g+G}$ in (C.62).  Including these and the above corrections,
the first two equations of (C.63) should read:
\bea H_{\mu\nu} &=& H^0_{\mu\nu} + H^g_{\mu\nu} -
x\(10 + 2x^2\rho_i\rho^i\)\lln\(F^a_{\mu\rho}F_{a\nu}^{\;\;\;\;\rho} - 
{1\over4}g_{\mu\nu}F^a_{\rho\sigma} F_a^{\rho\sigma}\), \nonumber \\ 
\epsilon_0 &=& \(\epsilon_0\)_0 + \epsilon^g_0 
- \lln\lbr{70\over 3}\D + 2x^2\rho_i\rho^i\D + {2\over 3x}\D_aD_i(T^az)^i\rbr, 
\nonumber \eea
and (C.64) should read:
\bea \Delta_r\L &=& \(\Delta_r\L\)_0 + \Delta_{rg}\L 
+ \lln\Bigg\{{N-99\over3}\D^2 - {2N+194\over3}\D\hV - 
{4N+32\over3}\D M_\psi^2 \nonumber \\ & &
+\(\D_\mu z^i\D^\mu\z^{\m}K_{i\m} -2V\)\[2x^2\rho_i\rho^i\D
+ {2\over 3x}\D_aD_i(T^az)^i \]\nonumber \\ & &
-2\D e^{-K}\(A_{ij}\A^{ij} - {2\over 3}R^i_jA_i\A^j\) 
+ {1\over3}\D\D_\mu z^i\D^\mu\z^{\m}\[4R_{i\m} - (N-57)K_{i\m}\]
\nonumber \\ & & + \({N+29\over6} - 2x^2\rho_i\rho^i\)\[2x^2\cW^{ab}\cbW_{ab}
+ \(\cW^{ab} + \cbW^{ab}\)\D_a\D_b + 2\D^2\]
\nonumber \\ & & + \({N+71\over3} - 2x^2\rho_i\rho^i\){x\over4}
F^a_{\rho\sigma}F_a^{\rho\sigma}\D_\mu z^i\D^\mu\z^{\m}K_{i\m}
\nonumber \\ & & - \({N+71\over3} - 2x^2\rho_i\rho^i\)x
F^a_{\rho\mu}F_a^{\rho\nu}\D_\nu z^i\D^\mu\z^{\m}K_{i\m}
\Bigg\}.\nonumber \eea 

\item The sign of the last term in the expression for $D^2 + H_{Gh}$ in (2.12)
of~\cite{us} and in (C.14) of~\cite{us2} is incorrect.  As a consequence, 
$-18\Gamma_{\mu\nu}\Gamma^{\mu\nu}$ in footnote 22 of~\cite{us2} and 
$-6\Gamma_{\mu\nu}\Gamma^{\mu\nu}$ in (B18) of~\cite{us} should both be
replaced by $-2\Gamma_{\mu\nu}\Gamma^{\mu\nu}$ in (C.61). 

\item In the expressions for $[D_\mu,D_\nu]$ for fermions, $ 
\Gamma_{\mu\nu}\to\Gamma_{\mu\nu} - {i\over2}F^a_{\mu\nu}D_a.$
As a consequence of this and the
above item, the coefficient $-24$ should be replaced by
$+2$ in $\Tr H^2_{Gh}$, Eq. (C.61) of~\cite{us2}, and the coefficient of
$\D_a\D^bF^a_{\mu\nu}F_b^{\mu\nu}$ should be ${1\over2}$ instead of 2 in the
same equation.  In addition the final results (4.6-8) and (5.2) of~\cite{us2}
are modified by the addition of the terms
$$ - {1\over3}\(N + 7 + N_G\)\[i\D^aF_a^{\mu\nu}\D_\mu z^iK_{i\m}\D_\nu\z^{\m} 
+ {1\over2}\D_a\D_b\(\cW^{ab} + \cbW^{ab}\) + 2\D^2\] $$ $$
+ {2\over3}\[i\D^aF_a^{\mu\nu}\D_\mu z^iR_{i\m}\D_\nu\z^{\m} +
D_i(T_az)^i\lbr\D_b\(\cW^{ab} + \cbW^{ab}\) + {2\over x}\D\D_a\rbr\] $$
from contributions proportional to $[D_\mu,D_\nu]^2$ from fermion loops
and ${1\over6}${\bf Tr}$G^2_{Gh}$, the term
$$ + 2x^2\rho^j\rho_j\[\D_a\D_b\(\cW^{ab} + \cbW^{ab}\) + 4\D^2\] $$ 
from $-{1\over4}\btr H_1^\chi H_3^\chi + t_\chi -{1\over4}\btr
H_1^gH_3^g+T_g$,  Eqs. (C.34,35,59)
of~\cite{us}, and an additional term 
$$ - 2\[\D_a\D_b\(\cW^{ab} + \cbW^{ab}\) + 4\D^2\] $$ 
from $-{1\over4}\btr H_1^gH_3^g$.  In addition the contribution of
$R_{\mu\nu}$ was neglected in the calculation of $2t_\chi$; this gives an
additional contribution 
$$ -2i\D_\mu z^k\[x\D_\nu\z^{\m}\rho_{\m jk} + \rho_{jk}\(\pp_\nu x -
i\pp_\nu y\)\]\[x\rho^j\D^aF_a^{\mu\nu} + 2(T_az)^j\(F_a^{\mu\nu} -
i\tF_a^{\mu\nu}\)\] $$
$$ + 2 \rho^j\rho_j\pp_\mu x \pp_\nu y\D^aF_a^{\mu\nu} + {\rm h.c.}, $$
which does not contribute to (2.22), and only the last term
contributes when the string dilaton is present.

\item The coefficient of
$\D_\mu z^i\D_\nu\z^{\m}K_{i\m}R_{j\n}\(\D^\mu z^j\D^\nu\z^{\n} - \D^\nu 
z^j\D^\mu\z^{\n}\)$ in footnotes 6, 13 and 21 and the coefficient of
$${1\over3}\D_\mu z^i\D_\nu\z^{\m}K_{i\m}\sum_\alpha\(N_\alpha+1\)K^\alpha_{j\n}
\(\D^\mu z^j\D^\nu\z^{\n} - \D^\nu z^j\D^\mu\z^{\n}\)$$
in footnote 8 of~\cite{us2} should be multiplied by $-2$.

\item The last term in brackets in the expression for $\btr(H_3^\chi)^2$
in (C.33)
of~\cite{us2} should be multiplied by ${1\over2}$, and the last term in (C.38) 
should be multiplied by $-2$, with corresponding changes in
(C.36) and the final results.

\item There are errors in the coefficients of the the expressions following 
$-T^{\chi g}_4$ in the second equality for ${1\over8}\btr\(H_1^{\chi g}\)^2$, 
Eq. (C.41), and in similar terms in the other traces. For the canonical gauge 
kinetic energy case considered here the corrections to amount to
the changes: $-2\D\hV - 6\D M^2$ in (C.41),
$-28\D M^2$ in the expressions for ${1\over2}\STr
H^2_\chi$, Eq.(C.36), $+ 8\D M^2$ and $-8\D M^2$ in 
${1\over8}\btr\(H_1^{\chi G}\)^2,$ and ${1\over2}\STr\(H_1^{\chi G}\)^2$,
respectively, Eqs. (C.50,51), and $+4\D M^2$
in ${1\over8}\btr\(H_1^{g+ G}\)^2,$ Eq.(C.58).

\item The following are misprints in~\cite{us2}: \newline 
The second line of (B.20) should be multiplied by $x^{-1}$.
\newline 
{\bf Tr}$\(\hG_\Theta^{\chi 
g}\)^2$ should be multiplied by ${1\over2}$ in the first line of (C.46); the
sign of the last term in footnote 23 is incorrect. \newline
The terms quartic in the field strength in (C.52--58) should be multiplied by
$x^2$. \newline
$(N + 5)/r^2 \to 5/r^2$ in (C.58). \newline
The terms $M^2_\lambda(\pp_\mu y\pp^\mu y/x^2)$ in (C.67,70) should be 
multiplied by 4. \newline
$M_\psi^2\to M_\psi^4$ in the second line of (C.71), and there should be a $+$
sign in front of the third from last line. \newline
In addition, a factor $\D_\mu\z^{\m}\D^\mu z^i$ is missing from the coefficient
of \newline $2K_{i\m}\(\hV + 2M^2_\psi\)$ in the expression for 
${1\over4}\Tr|D_\mu M_\theta|^2$ in (B12) of~\cite{us}.
\end{enumerate}  }}

\end{document}